\documentclass[journal=acsnano,manuscript=article]{achemso}
\usepackage{amsmath}

\author{Philippe Roelli}
\affiliation[San Sebastian]
{Nano-optics Group, CIC nanoGUNE BRTA, E-20018 Donostia-San Sebastián, Spain}
\author{Huatian Hu}
\affiliation[Lecce]
{Center for Biomolecular Nanotechnologies, Istituto Italiano di Tecnologia, via Barsanti 14, Arnesano, 73010, Italy}
\author{Ewold Verhagen}
\affiliation[AMOLF]
{Center for Nanophotonics, NWO Institute AMOLF, Science Park 104, 1098 XG Amsterdam, the Netherlands}
\author{Stephanie Reich}
\affiliation[FUB]
{Department of Physics, Freie Universität Berlin, 14195 Berlin, Germany}
\author{Christophe Galland}
\email{chris.galland@epfl.ch}
\affiliation[EPFL]
{Institute of Physics, Swiss Federal Institute of Technology Lausanne (EPFL), CH-1015 Lausanne, Switzerland \\ 
Center of Quantum Science and Engineering, Swiss Federal Institute of Technology Lausanne (EPFL), CH-1015 Lausanne, Switzerland}

\title{Nanocavities for Molecular Optomechanics:  their fundamental description and applications} 
\SectionNumbersOn

\begin{document}

\begin{tocentry}
\includegraphics[keepaspectratio, width=\textwidth]{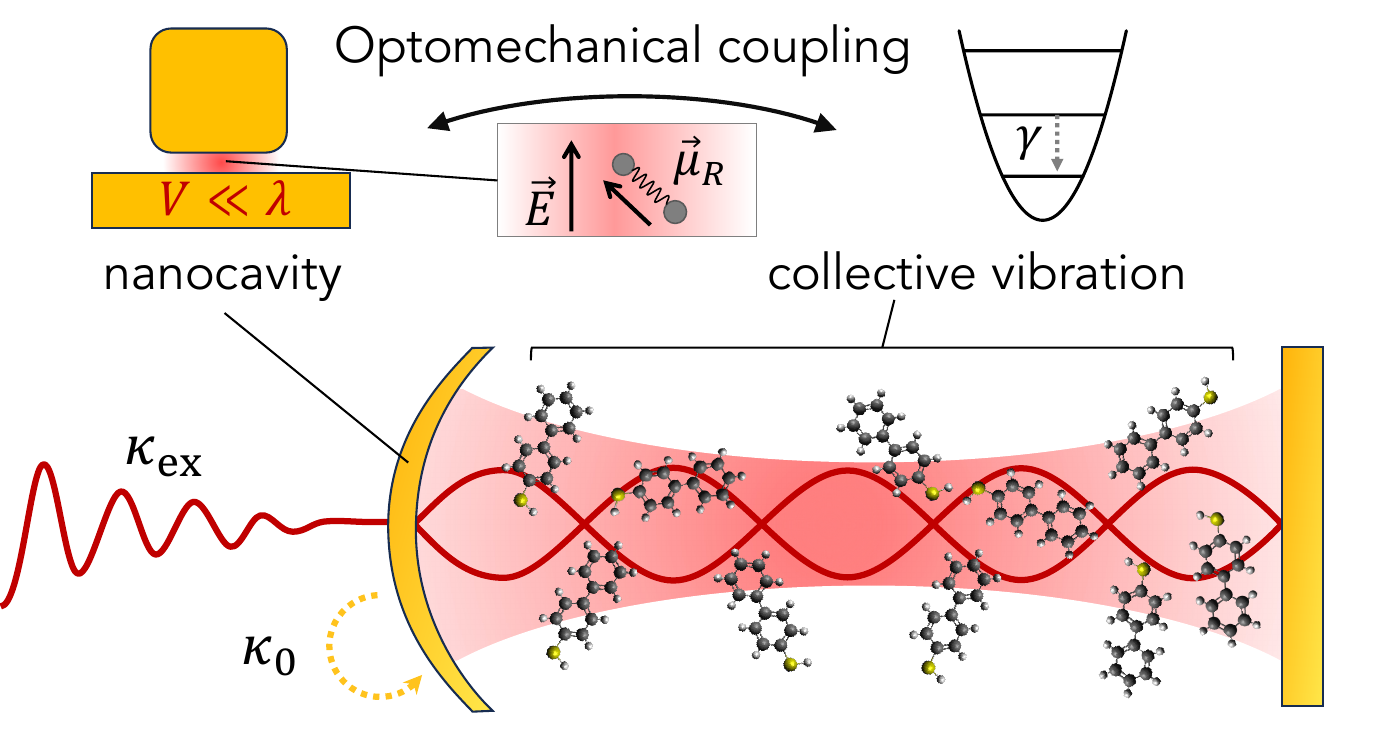}
\end{tocentry}

\renewcommand{\baselinestretch}{1.3}\small

\begin{abstract}
  \textbf{Vibrational Raman scattering -- a process where light exchanges energy with a molecular vibration through inelastic scattering -- is most fundamentally described in a quantum framework where both light and vibration are quantized. When the Raman scatterer is embedded inside a plasmonic nanocavity, as in some sufficiently controlled implementations of surface-enhanced Raman scattering (SERS), the coupled system realizes an optomechanical cavity, where coherent and parametrically amplified light-vibration interaction becomes a resource for vibrational state engineering and nanoscale nonlinear optics. 
  The purpose of this Perspective is to clarify the connection between the languages and parameters used in the fields of molecular cavity optomechanics (McOM) vs. its conventional, `macroscopic' counterpart, and to summarize the main results achieved so far in McOM and the most pressing experimental and theoretical challenges. We aim to make the theoretical framework of molecular cavity optomechanics practically usable for the SERS and nanoplasmonics community at large. 
  While quality factors ($Q$'s) and mode volumes ($V$'s) essentially describe the performance of a nanocavity in enhancing light-matter interaction, we point to the light-cavity coupling efficiencies ($\eta$'s) and optomechanical cooperativities ($\mathcal{C}$'s) as the key parameters for molecular optomechanics. As an illustration of the significance of these quantities, we investigate the feasibility of observing optomechanically induced transparency with a molecular vibration -- a measurement that would allow for a direct estimate of the optomechanical cooperativity.     }
\end{abstract}

\subsection*{Introduction}

Molecular cavity optomechanics (McOM) aims at unifying under a common theoretical description and a common language two distinct research areas: surface- and tip-enhanced Raman scattering (SERS and TERS) on the one hand, and  (macroscopic) cavity optomechanics on the other hand. 
 SERS and TERS were mostly developed by chemists, spectroscopists, and surface scientists; they describe the huge enhancement of vibrational Raman scattering intensity from molecules interacting with the near field of metallic nanostructures supporting localised surface plasmon polaritons. \cite{ru2008} 
In contrast, cavity optomechanics studies the coherent interaction (down to the quantum regime) between light trapped in a dielectric cavity and a mechanical oscillator \cite{aspelmeyer2014} whose mass may range from an atomically thin membrane \cite{xu2022a} to a kilogram-scale suspended mirror \cite{whittle2021}. 

First released in Ref.~\citenum{roelli2016} and initially motivated by the experimental observation of anomalously narrow SERS enhancement profiles under laser wavelength scans in Ref.~\citenum{zhu2014}, this connection prompted the development of full quantum models for SERS and TERS that account for the different forms of backaction imparted on the molecular vibration by the laser-driven plasmonic resonance \cite{roelli2016,schmidt2016a,schmidt2017,kamandardezfouli2022,esteban2022,bourgeois2024}. 
 McOM has been fuelling inspiration for new conceptual and experimental developments in the field of SERS and TERS, where the focus is not on chemical or material analysis, but rather on achieving classical and quantum control of molecular vibrations and developing new nanoscale optical devices leveraging optomechanical nonlinearities \cite{chen2021}.
Progress in McOM hinges on two pillars: (i) a proper understanding of the theoretical framework and its connection with designed and measured nanocavity parameters, and (ii) an improved control and power resilience of device parameters governing light-nanocavity and nanocavity-vibration couplings, supplemented with critical attention to other sources of nonlinearities that may interfere with optomechanical phenomena. 
Mastering these two pillars will allow the engineering of new nanocavities and molecules that feature improved stability and finer control over the plasmon-molecule coupling and dissipation rates, with the overarching goal of reaching large optomechanical cooperativities ($\mathcal{C}$) and nanocavity coupling efficiencies 
close to unity -- which has become routine for micro-fabricated oscillators coupled to dielectric cavities \cite{aspelmeyer2014}. 

These two pillars are discussed in this Perspective after a brief reminder of the conventional theoretical description of SERS. In this Perspective, we place special emphasis on the description of light-nanocavity coupling and aim to clarify its connection with the input-output formalism used in cavity optomechanics. We will therefore present a more complete analogy between a typical SERS experiment on a plasmonic nanocavity and a macroscopic cavity optomechanical setup, highlighting the specificities of McOM. We also propose a scheme to perform optomechanically induced transparency measurement in McOM, a technique that would provide direct access to the optomechanical cooperativity. We conclude with an overview of some of the most appealing perspectives in the field. 

\subsection*{Basics of vibrational Raman scattering and SERS enhancement mechanisms}

Before describing the theoretical and experimental aspects of McOM, we briefly review the broadly accepted concepts underlying SERS and TERS. The Raman effect \cite{ru2008} is first described in a classical model: a molecule under the applied monochromatic field $\mathbf{E}$ oscillating at frequency $\omega_L$ will experience an induced dipole $\boldsymbol{\mu}(t)=\boldsymbol{\alpha}(t)\cdot \mathbf{E} \cos{\omega_L t}$ where $\boldsymbol{\alpha}(t)\simeq \boldsymbol{\alpha}_0+\boldsymbol{\alpha}_R \cos{\Omega_\nu t}$ is the 3-dimensional second rank polarizability tensor that is modulated by the vibration. 
For simplicity, we will consider here only a single normal mode with vibrational frequency $\Omega_\nu \ll \omega_L$ (vibrational frequencies are in the range of 1-100 THz while laser frequencies are typically above 500 THz) and associated normal mode coordinate $Q_\nu$. This mode is Raman active if $\boldsymbol{\alpha}_R\propto\frac{\partial \boldsymbol{\alpha}}{\partial Q_\nu}\neq \boldsymbol{0}$; in this case, for appropriate orientation of the incoming field, the Raman dipole $\boldsymbol{\mu}_R(t)= \frac{1}{2} \boldsymbol{\alpha}_R\cdot\mathbf{E} \left\{ \cos{(\omega_L-\Omega_\nu) t} + \cos{(\omega_L+\Omega_\nu) t} \right\}$ radiates at two new frequencies $\omega_{S}=\omega_L-\Omega_\nu$ (Stokes) and
$\omega_{aS}=\omega_L+\Omega_\nu$ (anti-Stokes), which constitute the inelastic Raman scattered field. Note that the classical model predicts Stokes and anti-Stokes sidebands of equal amplitudes; a semi-classical model where the vibration is quantized correctly predicts the observed asymmetry \cite{loudon1997,ru2008}.

SERS was discovered in the 1970s as molecules absorbed on metal electrodes and other rough metal surfaces were studied with Raman spectroscopy \cite{fleischmann1974,albrecht1977,jeanmaire1977}. The scattering intensity from molecules showed a great enhancement, which was soon realized to be a genuine increase in the scattering cross section per molecule and not simply due to the aggregation of molecules on the metal surface.
The main enhancement mechanism is called electromagnetic enhancement, arising from the optical near fields of localized surface plasmons close to a metal nanostructure. \cite{ru2008,moskovits2013,ding2017} 
A second enhancement mechanism arises from the chemical interaction between molecule and metal. For example, charge transfer may change the electronic configuration and even vibrational frequencies of the molecules and affect their Raman cross sections. Within McOM, the chemical enhancement has to be included ad hoc in the value of $\boldsymbol{\alpha}_R$ for the molecules. We will thus not discuss it further below, even though it impacts experiments where molecules are directly bound to the metal \cite{morton2009,valley2013,chen2023a}.

When the near field of a localized surface plasmon exceeds the incident field at frequency $\omega$ by an enhancement factor $K(\omega)$, it can be shown that the probability of generating the Raman dipole is enhanced by $K^2(\omega_L)$ and that of radiating the scattered light into the far-field by approximately $K^2(\omega_S)$ \cite{leru2006a}. For a single molecule, the SERS intensity compared to that of Raman scattering in free space is therefore multiplied by the electromagnetic enhancement factor 
\begin{equation*}
  K^2(\omega_L)K^2(\omega_S)\simeq K^4(\omega_L).
\end{equation*}
where the last approximation is valid when $\Omega_\nu$ is not too large compared to the plasmonic resonance bandwidth (corresponding to the unresolved sideband regime of McOM).

The strength of this approach is that the enhancement factor $K(\omega)$ can be calculated by solving Maxwell's equations only. For well-defined nanostructures with voids or cavities of 1-10\,nm dimension, a typical near-field enhancement is $K\approx 10^2$, resulting in $\sim 10^8$ electromagnetic SERS enhancement factors for a probe inside the hot spot. The spatial distribution of plasmonic hotspots, as well as the dependence of the enhancement on polarization and scatterer orientation are predicted exceptionally well within this framework. \cite{kusch2017} However, it has been pointed out that this approach may significantly underestimate the absolute enhancement factors and overestimate the SERS resonance bandwidth in some experiments.\cite{heeg2021} 

In the microscopic, semi-classical theory of Raman scattering, SERS is viewed as a higher-order Raman process (HoRa),\cite{mueller2019} which is particularly instructive as a similar point of view is taken in McOM. The microscopic theory treats the normal Raman effect within perturbation theory, where a molecule is first excited from its ground state to an excited electronic state and reaches back to the electronic ground state but in an excited vibrational state. The difference between the two transition energies is equal to the energy of the vibration $\hbar\Omega_\nu$. SERS is modelled by adding two interaction steps that describe molecular excitation and emission through the localized surface plasmon of resonance frequency $\omega_{p}$ and linewidth $\kappa$. 
Two resonances occur when the incoming $\hbar\omega_L$ or scattered $\hbar\omega_s=\hbar\omega_L-\hbar\Omega_\nu$ photon energy matches the plasmon energy. 
A coupling factor describes the plasmon-related interaction in the SERS process, as detailed in Ref.~\citenum{mueller2019}. 
Note that this framework does not keep track of possible changes in vibrational population and dynamics that could be induced by Raman scattering, which is a gap that the McOM framework fills. 

\subsection*{Optomechanical model of nanocavity-enhanced Raman scattering}

\begin{figure}
    \centering
    \includegraphics[keepaspectratio, width=0.6\textwidth]{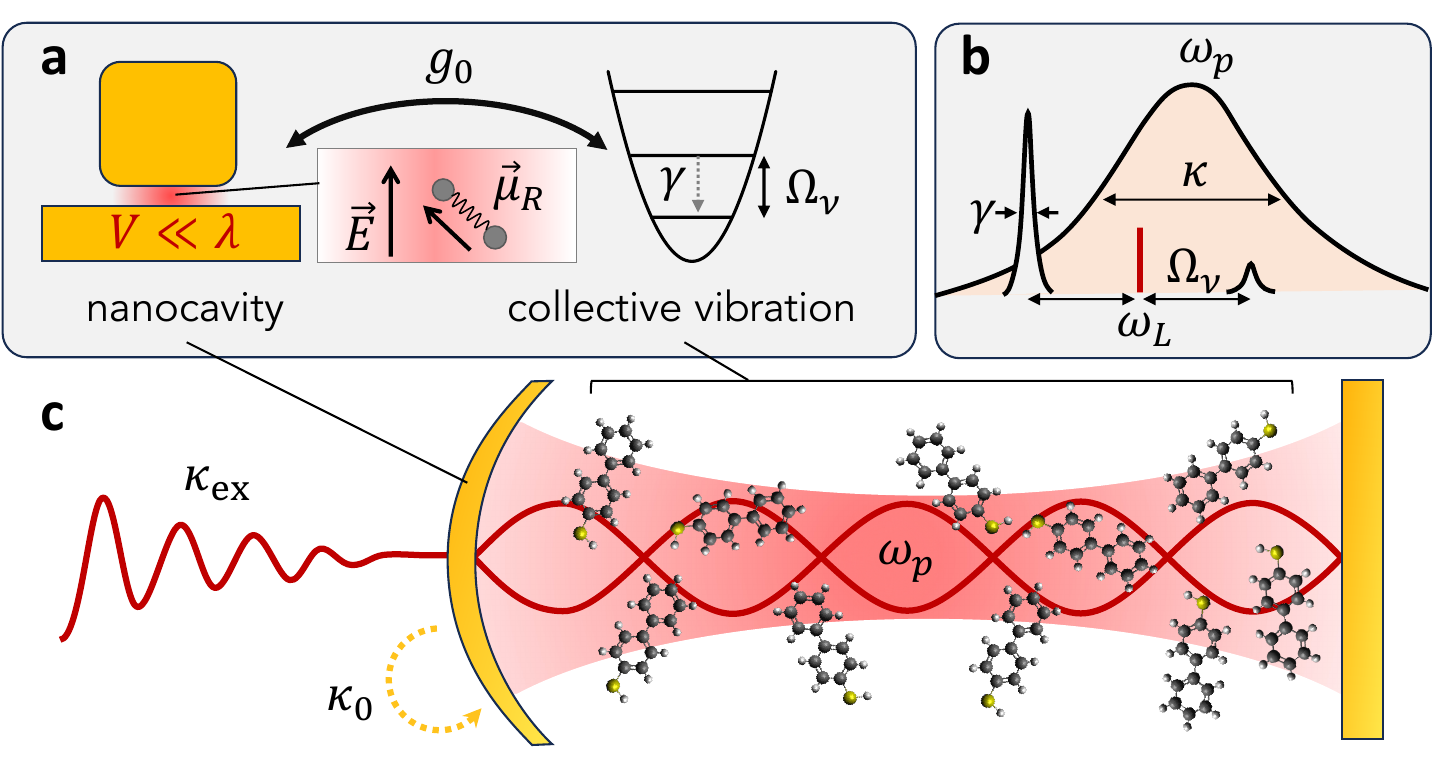}
    \caption{\footnotesize Simplified setting for single-mode molecular cavity optomechanics (McOM). (a) A plasmonic cavity (here sketch as a metallic dimer) supports a plasmonic resonance with radiative and non-radiative decay rates $\kappa_{\rm ex}$ and $\kappa_0$, respectively. The surface-enhanced local field $\boldsymbol{E}(t)$ couples to the induced Raman dipole $\boldsymbol{\mu}_R(t)$ of the vibrating molecule, resulting in the vacuum optomechanical coupling rate $g_0$. The molecular vibration is damped at a rate $\gamma$. (b) Frequency domain schematic showing the plasmonic resonance at frequency $\omega_p$ of width $\kappa=\kappa_{\rm ex}+\kappa_0$. Under a single-frequency pump at $\omega_L$ the optomechanical coupling gives rise to two Raman sidebands at $\omega_L\pm\Omega_\nu$, where $\Omega_\nu$ is the molecular vibration frequency.
    Panel (c) is inspired from Ref.~\citenum{sauvan2013}. 
    }
    \label{fig:intro}
\end{figure}

Molecular cavity optomechanics (McOM) is based on a quantum model of plasmon-enhanced Raman scattering; in a minimal setting one molecular vibrational mode, treated as a harmonic oscillator with resonance frequency $\Omega_\nu$ and decay rate $\gamma$, modulates the optical response of the nanocavity characterized by one plasmonic mode with resonance $\omega_p \gg \Omega_\nu$ and total decay rate $\kappa \gg \gamma$ (this dissipation hierarchy is reversed in Raman lasers based on high-$Q$ dielectric cavities \cite{latawiec2018} where $\kappa \ll \gamma$, the rest of the physics being otherwise similar), see Fig.~\ref{fig:intro}. 
Both vibrational and plasmonic degrees-of-freedom obey boson statistics and are supposed to be exchanging energy quanta at rates $\kappa$ and $\gamma$ with a thermal bath. 
The resolved-sideband regime is achieved when $\Omega_\nu > \kappa/2$ (Fig.~\ref{fig:intro}b), which is possible in McOM despite the fast plasmon decay thanks to the often very high vibrational frequencies (tens of THz). For example, for a $Q$-factor of 10 at 700~nm, the minimum vibrational frequency to be sideband-resolved is $715$~cm$^{-1}$ or 21.4~THz. 

Pure dephasing is not observed to play a significant role in plasmonic decoherence \cite{sonnichsen2002}. For vibrational modes of molecules, experimental data obtained using 2D IR spectroscopy on monolayers suggest that inhomogeneous broadening strictly dominates over population relaxation and single-molecule pure dephasing rates. \cite{kraack2015,yan2016,metzger2019,triana2022,wilcken2023} For configurations typically studied in SERS, the inhomogeneous (collective) relaxation rate was recently measured to be sub-picosecond \cite{jakob2024}, consistent with earlier vibrational sum-frequency studies on monolayers \cite{bordenyuk2005,stiopkin2008}. In the single-molecule limit, a report from 2014 inferred much longer-lived coherence between two vibrational modes of the same molecule, in excess of 10~ps. \cite{yampolsky2014}  If such a result were to be confirmed, it would imply that single-molecule optomechanics, as achieved in the regime of picocavities \cite{benz2016} or in TERS \cite{zhang2013a}, may benefit from reduced vibrational decay rate $\gamma$ and a corresponding increase in the cooperativity (introduced below).


The single-photon optomechanical coupling rate $g_0$ is proportional to $\left( \mathbf{u}^*\cdot\frac{\partial \boldsymbol{\alpha}}{\partial Q_\nu}\cdot\mathbf{u} \right)/V$ with $V$ the plasmonic mode volume and $\mathbf{u}$ the near field polarisation vector. Note that the term in parenthesis is proportional to the square root of the Raman cross-section of the molecule \cite{roelli2016}. 
There are two main interpretations for the meaning of $g_0$: (i) Conventionally, $g_0/2\pi$ measures the plasmonic resonance frequency shift induced by a vibrational mode displacement whose magnitude equals that of the ground state fluctuations (or zero-point motion) \cite{aspelmeyer2014, roelli2016}; (ii) Alternatively, from a microscopic point of view the interaction energy $\hbar g_0$ relates to the interaction energy $\frac{1}{2}\boldsymbol{\mu}_R\cdot \mathbf{E}$ between the local vacuum electric field $\mathbf{E}$ and the induced Raman dipole $\boldsymbol{\mu}_R$ (Fig.~\ref{fig:intro}a) \cite{schmidt2017}.

For weak single-photon optomechanical coupling ($g_0\ll\kappa$, as is the case in all realizations of McOM to date) the interaction Hamiltonian can be linearized. 
The vibrational and plasmonic modes coherently exchange quanta at a rate $\Omega=2 g_0\sqrt{n_p\,N}$, where $N$ is the effective number of coupled molecules and $n_p$ the time-averaged occupancy of the plasmonic mode (intracavity plasmon number), which depends on the input laser power $P_L$ and the overall incoupling efficiency and incoupling rate (see below). As a result, for weak coupling $\Omega\ll\kappa$, the mechanical vibrations are coupled to the optical bath to which the plasmon mode decays at rate $\Gamma_\text{opt} \propto 4 n_p  N g_0^2/\kappa$, scaling linearly with both the number of molecules and the number of intracavity plasmons. 
As the single most important parameter characterising the performance of an optomechanical cavity, the cooperativity is defined as $\mathcal{C}=\Gamma_\text{opt}/\gamma=4 n_p N g_0^2/(\kappa\gamma) = n_p \mathcal{C}_0$, which measures the ratio of $\Gamma_\text{opt}$ to the decay rate of the vibration. Here, $\mathcal{C}_0=4 N g_0^2/(\kappa\gamma)$ is known as the single-photon cooperativity. 

\begin{figure}
    \centering
    \includegraphics[keepaspectratio, width=\textwidth]{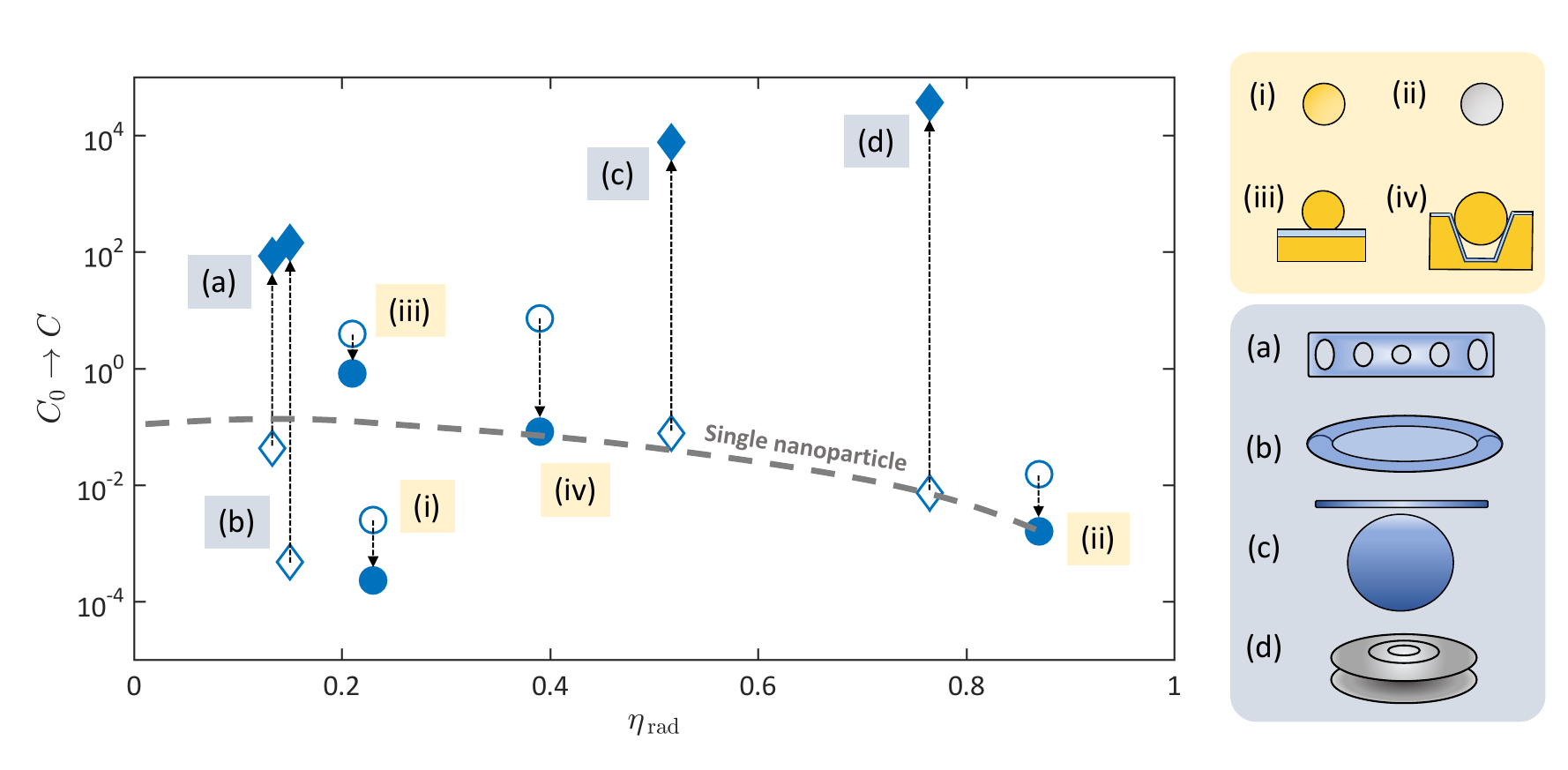}
    \caption{\footnotesize Overview of representative plasmonic (i to iv) and dielectric (a to d) optomechanical cavities in terms of cooperativity and radiative coupling efficiency (defined in Fig.~\ref{fig:etas}). For plasmonic cavities, we assume that a monolayer of biphenyl-thiol molecules covers the nanoparticle or the substrate and the incoming laser power is set as 100~$\mu$W in a diffraction limited spot. Empty symbols represent the single-photon cooperativity $\mathcal{C}_0$, while full blue symbols represent $\mathcal{C}={n}_p \mathcal{C}_0$. 
    The data for dielectric cavities are compiled from published experiments: 
    (a) Ref.~\citenum{chan2011}; 
    (b) Ref.~\citenum{verhagen2012}; 
    (c) Ref.~\citenum{wilson2015}; 
    (d) Ref.~\citenum{teufel2011}.
    }
    \label{fig:overview}
\end{figure}

\begin{figure}
    \centering
    \includegraphics[keepaspectratio, width=0.7\textwidth]{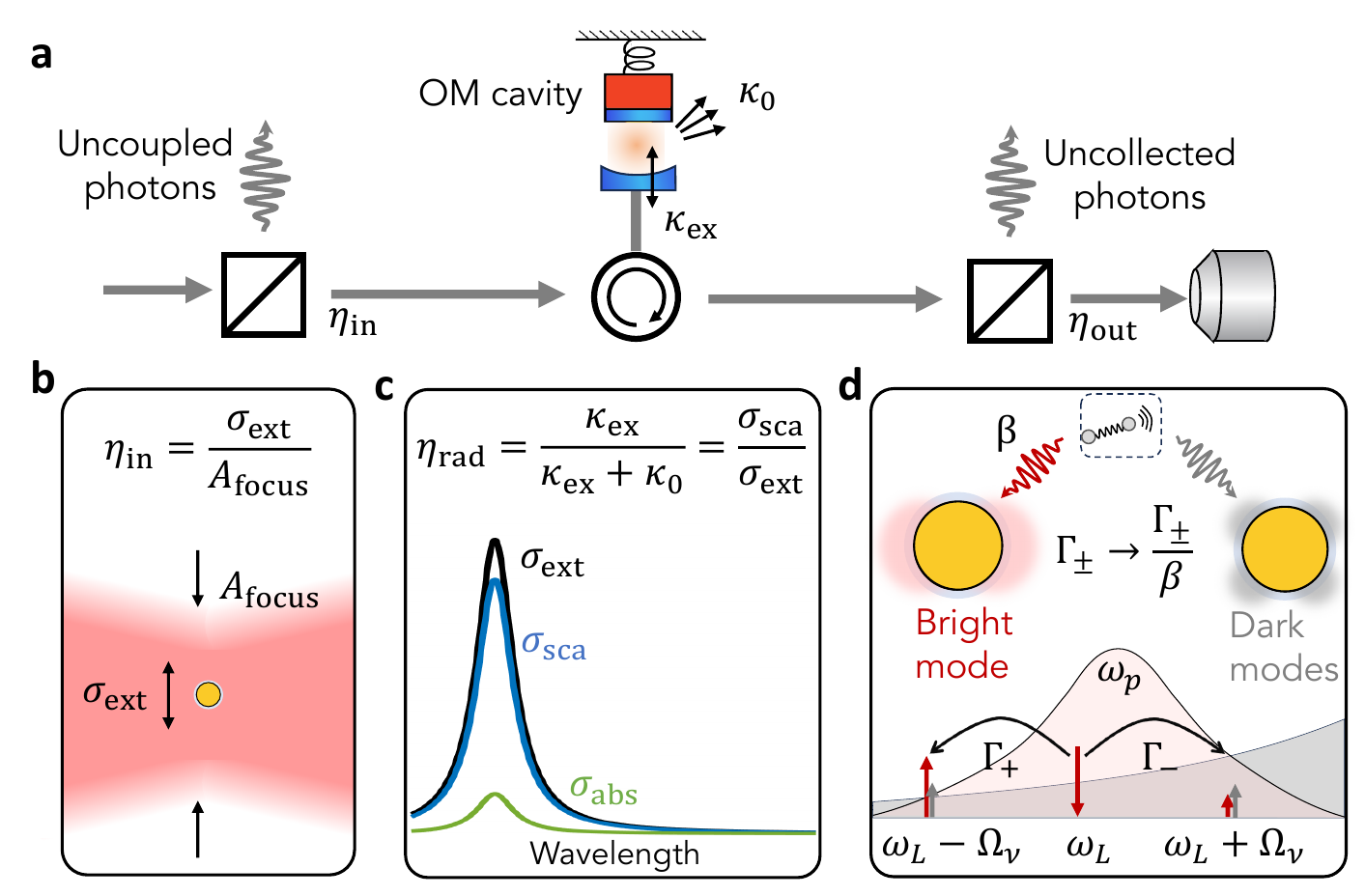}
    \caption{\footnotesize Correspondence between macroscopic and molecular cavity optomechanics. (a) Sketch of general cavity optomechanics scenario where all input and output coupling losses can be modelled by beam-splitters with transmissions $\eta_\text{in}, \eta_\text{out}$, respectively; the total decay rate of the cavity is the sum of an external decay rate $\kappa_\text{ex}$ allowing for photon exchange with radiation modes, and an internal decay rate $\kappa_0$ that accounts for absorption and scattering losses. (b) In McOM the input coupling efficiency corresponds to the ratio of focal spot size to the extinction cross-section of the nanocavity (pictured here as a simple metal nanoparticle coated with molecules). (c) The external coupling efficiency is given by the ratio of scattering to extinction cross-section in the single mode limit. (d) The optomechanical coupling occurs in the near-field through the Raman polarizability, leading to scattering rates $\Gamma_{+,-}$ for Stokes (add a quantum of vibration), resp. anti-Stokes (removes one) processes. The existence of a spectrally overlapping quasi-continuum of dark modes is responsible for additional Raman scattering contributions in the near-field that are however not detected in the far-field (quenching). 
    }
    \label{fig:etas}
\end{figure}

Achieving $\mathcal{C}\geq 1$ is a prerequisite for realising key applications of cavity optomechanics. As a first example, in the resolved-sideband limit, a laser drive optimally blue-detuned from the cavity resonance modifies the damping rate of the vibration to $\tilde{\gamma} \simeq \gamma (1-\mathcal{C})$, causing a linewidth reduction and coherent amplitude amplification of the vibrational mode \cite{aspelmeyer2014} (cold damping, or cooling, is possible under red-detuned drive). More generally, $\mathcal{C}\sim 1$ marks the onset of strongly nonlinear dynamics.  
As a second example, when using an optomechanical system to implement coherent optical frequency conversion \cite{han2021} the efficiency scales as $\mathcal{C}\mathcal{C}'/(1+\mathcal{C}+\mathcal{C}')^2$, where $\mathcal{C}'$ represents the cooperativity computed for the other cavity resonance involved. A necessary condition for optimal conversion is therefore $\mathcal{C}=\mathcal{C}'\gg 1$. 
Figure~\ref{fig:overview} presents how different nanocavity geometries compare with typical dielectric cavities used in optomechanics in terms of cooperativity and radiative coupling efficiency $\eta_\text{rad}$, the latter being a mode-specific property of the nanocavity quantifying its ability to couple to far-field radiation (see definition in Fig.~\ref{fig:etas}c).  Here one recognizes that, while McOM nanocavities can naturally feature single-photon cooperativities that exceed those of state-of-art dielectric optomechanical resonators, enhancing the coupling through large cavity occupancies is more challenging.

The McOM theory was not initially proposed as a new fundamental form of plasmon-vibration interaction nor a substitute to the accepted SERS enhancement\cite{leru2024} mechanisms reviewed above; it was shown to recover the same predictions as the electromagnetic theory based on field enhancement in the regime where $\mathcal{C}\ll 1$. \cite{roelli2016}
However, a modified and nonlinear response of the cavity and molecular vibration is predicted when $\mathcal{C}$ approaches unity. Very strong pumping is required for reaching appreciable values of $n_p$ (recall that $\mathcal{C}= n_p \mathcal{C}_0$), under which nanocavities can be irreversibly damaged or display other nonlinearities, challenging the experimental realisation of this regime. \cite{lombardi2018,crampton2018,liu2021,xu2022,lai2022,jakob2023} 
At the same time, since the thermal phonon occupancy $n_{\rm th}$ is very low due to the high vibrational frequencies, the cooperativity is the same as the quantum cooperativity (defined as $\mathcal{C}/(1+n_{\rm th})$) in McOM systems, in contrast to most other optomechanical cavities (especially at room temperature) where the quantum cooperativity is much smaller. Moreover, McOM is a quantum formalism that treats vibrational and plasmonic degrees of freedom at the same level of description, with some predictions that go beyond the conventional theory of SERS.
Here is a non-exhaustive list of physical phenomena and observables that the McOM formalism predicts: \cite{esteban2022}

\begin{itemize}
    \item In the sideband-resolved regime ($\kappa<2\,\Omega_\nu$) and for blue-detuned pumping, a new enhancement mechanism may be achieved through dynamical backaction \cite{roelli2016}, akin to phonon lasing.
    \item Quanta of molecular vibration mediate quantum correlations between Raman scattered fields \cite{schmidt2021} -- as already observed in the absence of a nanocavity \cite{bustard2015,kasperczyk2015,parra-murillo2016,saraiva2017,anderson2018};
    \item The collective nature of SERS is important even in the spontaneous scattering regime, in the sense that the Raman scattered field is correlated with a collective quantum of vibration that is coherently shared among all molecules coupled to the nanocavity field \cite{roelli2016} -- an effect also observable in the absence of nanocavity through spatial mode filtering \cite{vento2023}. Cooperative effects among vibrating molecules are expected under sufficiently large cooperativity \cite{zhang2020};
    \item Vibrational pumping \cite{kneipp1996,haslett2000,leru2006,maher2006,maher2008,galloway2009,kozich2010,benz2016,long2016,crampton2018,arzumanyan2021,leru2024} can be understood as the manifestation of quantum backaction \cite{benz2016}, otherwise difficult to observe in macroscopic cavity optomechanics. It occurs also for resonant cavity pumping and in the unresolved-sideband regime, as soon as $\mathcal{C}\sim n_{\rm th}$. Note that charge transfer between the metal and the molecule was also proposed as a mechanism for vibrational pumping \cite{boerigter2016,shin2023,stefancu2024}, a process beyond the scope of McOM. 
    \item McOM provides a rigorous framework to study how the multimode nature of plasmonic or hybrid plasmonic-dielectric \cite{shlesinger2021,shlesinger2023} resonators impacts the SERS signal \cite{kamandardezfouli2017,zhang2021}. For example, multimode McOM predicts a much larger optical spring effect (shift of the Raman peak) with respect to the optomechanical damping rate (change in its linewidth), which results from the large real part of the dyadic Green's function in plasmonic nanocavities, itself a consequence of the quasi-continuum of plasmonic modes supported at shorter wavelengths in metallic nanogaps \cite{zhang2021};
    \item Vibrational sum- and difference-frequency generation can be computed with a proper account of quantum and backaction noise \cite{roelli2020}, with applications in coherent mid-infrared and THz frequency conversion \cite{roelli2020,chen2021} and in vibrational spectroscopy \cite{okuno2015,humbert2019,linke2019,xomalis2021,tanaka2022,lin2024}. 
\end{itemize}

\subsection*{Scaling laws}

The performance of a nanocavity in enhancing light-matter interaction in its near field is well characterized by the quality factor $Q$ and mode volume $V$ (which are modal quantities) \cite{wu2021}. For a single molecule acting as mechanical oscillator in McOM, it has been shown that the single-photon cooperativity $\mathcal{C}_0$ scales as $Q/V^2$ (instead of $Q/V$ for the Purcell factor).\cite{roelli2016}  It is because the field intensity enhancement factor $K^2$ introduced above scales as $1/V$ and the SERS intensity scales as $K^4$. In contrast with cavity QED, the OM coupling rate is field-enhanced: If we account for the fact that the intracavity plasmon occupancy $n_p$ scales with $Q$ (if the coupling efficiency remains fixed) we find that $\mathcal{C}$ scales as $(Q/V)^2$ for a single molecule. In practice, however, $n_p$ is often limited by material failure and optically induced instabilities.

Most experiments deal with ensembles of $N$ molecules collectively coupled to the nanocavity mode, resulting in a factor $N$ increase in $\mathcal{C}_0$. \cite{roelli2016,zhang2020} In the limit where molecules fill the entire mode volume, i.e. when $N\propto V$, we obtain the scaling $\mathcal{C}_0\propto Q/V$. As a side note, for a resonant cavity-emitter interaction like in infrared absorption, the collective cooperativity is independent of mode volume, it only depends on the emitter concentration per unit volume. \cite{roelli2020} 

Unless the single-photon strong coupling regime is reached ($g_0\sim \kappa$), the parameter of relevance for McOM is not so much $\mathcal{C}_0$ as $\mathcal{C}= n_p\, \mathcal{C}_0$; we must therefore precisely estimate the intracavity plasmon number $n_p$, which depends on the far-field to near-field coupling efficiency \cite{baumberg2019,bedingfield2022,vento2023a} and on the decomposition of the total plasmon decay rate $\kappa$ into radiative and non-radiative channels \cite{faggiani2015,yang2015,marquier2017}, as is discussed in the next section. The various decay channels are also responsible for the existence of both dispersive and dissipative optomechanical coupling mechanisms, the latter contribution being much less understood while possibly dominant in some McOM realisations \cite{primo2020} (see below). We remark in passing that if the intracavity plasmon number $n_p$ could be made to increase linearly with mode volume, i.e. the electromagnetic energy density inside the nanocavity would be kept constant, then $\mathcal{C}$ would become independent on mode volume. But in practice the highest tolerable laser power is not constrained by mode volume alone and reducing the mode volume is a key pursuit to increase $\mathcal{C}$ in McOM.

\subsection*{Input-output formalism}

In order to take advantage of the vast body of knowledge in macroscopic cavity optomechanics \cite{aspelmeyer2014} to design and interpret McOM experiments, and to gain a physical intuition, we advocate for a proper use of the input-output formalism \cite{gardiner2004} in the context of plasmonic nanoantennas and nanocavities \cite{hamam2007}. 
Figure \ref{fig:etas} proposes a one-to-one correspondence between the canonical optomechanical cavity framework and that of McOM. We will analyse the situation in terms of photon flux entering and leaving the cavity, from the laser source to the detector. First, if we call $\Phi_L$ the photon flux impinging from the laser on the nano-antenna (here pictured as a simple metallic nanoparticle for simplicity), only a fraction $\Phi_\text{in} = \eta_\text{in}\,\Phi_L$ will actually interact with it. The factor $\eta_\text{in}$ quantifies the input coupling efficiency: in macroscopic cOM it can account for fiber-to-waveguide coupling losses, for example; in McOM it typically accounts for the overlap of the incident wavefront and the time-reversed radiation field of the nano-antenna, and thereby for the ratio of the extinction cross-section of the nano-antenna $\sigma_\text{ext}$ to the laser spot size $A_\text{focus}$ (Fig.~\ref{fig:etas}b). Note that for elongated nanorods, nanoparticle-on-mirror cavities and all other structures supporting several modes, the polarisation of the incoming field at the position of the nanocavity must be taken into account to estimate $\eta_\text{in}$ \cite{vento2023a}. A particularity of nanocavities is the difficulty of achieving $\eta_\text{in}\sim 1$, compared to macroscopic dielectric cavities.


In general, the input mode is populated by a coherent laser field plus the irreducible quantum fluctuations (shot noise); its amplitude couples to the cavity at rate $\sqrt{\kappa_\text{ex}}$, which for a simple Fabry-Perot cavity is controlled by the transmission of the input mirror and the round-trip time in the cavity. For a plasmonic nanocavity, $\kappa_\text{ex}$ measures the rate of radiative energy decay into the far field over the entire solid angle. By reciprocity, $\kappa_\text{ex}$ is therefore a measure of how efficiently an external field may excite a particular nanocavity mode (the subscript `ex' stands for `external'). 
Beside this useful cavity loss channel (in the sense that it connects the cavity to the experimentally controlled fields) other loss channels are captured by $\kappa_0$, including absorption in the metal or excitation of propagating surface plasmon polaritons eventually damped in the metal (for nanocavities built on metallic films). 
The equations of motion therefore include the vacuum quantum noise entering the cavity through these undesired channels, maintaining the fundamental connection between dissipation and fluctuation.

The total decay rate of the cavity is $\kappa=\kappa_\text{ex}+\kappa_0$, related to the quality factor by $Q=\omega_p/\kappa$. The quality factor alone hides the information on the relative magnitudes of $\kappa_\text{ex}$ and $\kappa_0$. For most applications in cavity optomechanics, it is beneficial to work with a critically or over-coupled nanocavity mode, i.e. having $\kappa_\text{ex}\geq \kappa_0$. If we define the external coupling efficiency of a nanocavity mode as $\eta_\text{rad}=\kappa_\text{ex}/\kappa$, then the aim is to reach $\eta_\text{rad}\geq 0.5$. 
For example, the efficiency of optomechanical frequency conversion is proportional to the products of $\eta_\text{rad}$'s at the two frequencies involved, so each $\eta_\text{rad}$ must be made close to unity. \cite{han2021} 
In terms of the scattering cross section $\sigma_\text{sca}$ and extinction cross section $\sigma_\text{ext}$ that are usually computed or measured for plasmonic antennas, we simply have $\eta_\text{rad}=\sigma_\text{sca}/\sigma_\text{ext}$, i.e., the nanoantenna albedo (see Fig.~\ref{fig:etas}c). In other words, a good nanocavity for McOM should have an extinction cross section that is dominated by scattering. We finally note that for several applications it will be important to maximally collect the scattered field on a detector. The collected fraction of the scattered power is denoted by $\eta_\text{out}$. 

A subtlety that is probably unique to McOM is the existence of a quasi-continuum of dark plasmonic modes that may spectrally overlap with the bright mode of interest (for which all parameters above are computed). Sometime referred to as the `pseudomode' in the regime of strong coupling \cite{delga2014,li2016}, this quasi-continuum is responsible for quenching the emission of a dipole placed in the near field; the $\beta$-factor is defined as the fraction of photons injected in the bright mode of interest vs. the total photon emission rate in the near-field that includes all quenching channels. \cite{wu2021} We explain how to compute  $\beta$ (which is a priori different for Stokes and anti-Stokes sidebands) in the Appendix. We recall that the $\beta$-factor is a useful figure-of-merit for single-photon sources embedded in photonic structures such as waveguides and micropillars, where it quantifies the likelihood that a photon is emitted in the guided mode vs. the continuum of free-space modes and is intimately linked to the Purcell factor. \cite{thyrrestrup2010}

The subtlety is that, in McOM, quenching by the quasi-continuum actually enhances the Raman scattering rate in the near field compared to the rate computed from the single-mode approximation presented above, by a factor $\frac{1}{\beta}$;  but the corresponding Raman photons are never detected (see Fig.~\ref{fig:etas}d). This `dark channel' for Raman scattering was invoked in recent observations of pronounced vibrational pumping while the detected Stokes scattering was much too low to compete with the expected vibrational relaxation rate. \cite{crampton2018}. 
When computing $g_0$ based on the full dyadic Green's function calculation, however, there is no need to correct the resulting Raman scattering rate by $\frac{1}{\beta}$ since that approach already accounts for the total photonic local density of states at the position of the molecule. \cite{kamandardezfouli2017,zhang2021} In that case, the collection efficiency -- including quenching effects -- is obtained from the computation of the Green's function between the molecule and the detector position. \cite{kamandardezfouli2017} 

\begin{figure}
    \centering
    \includegraphics[keepaspectratio, width=0.9\textwidth]{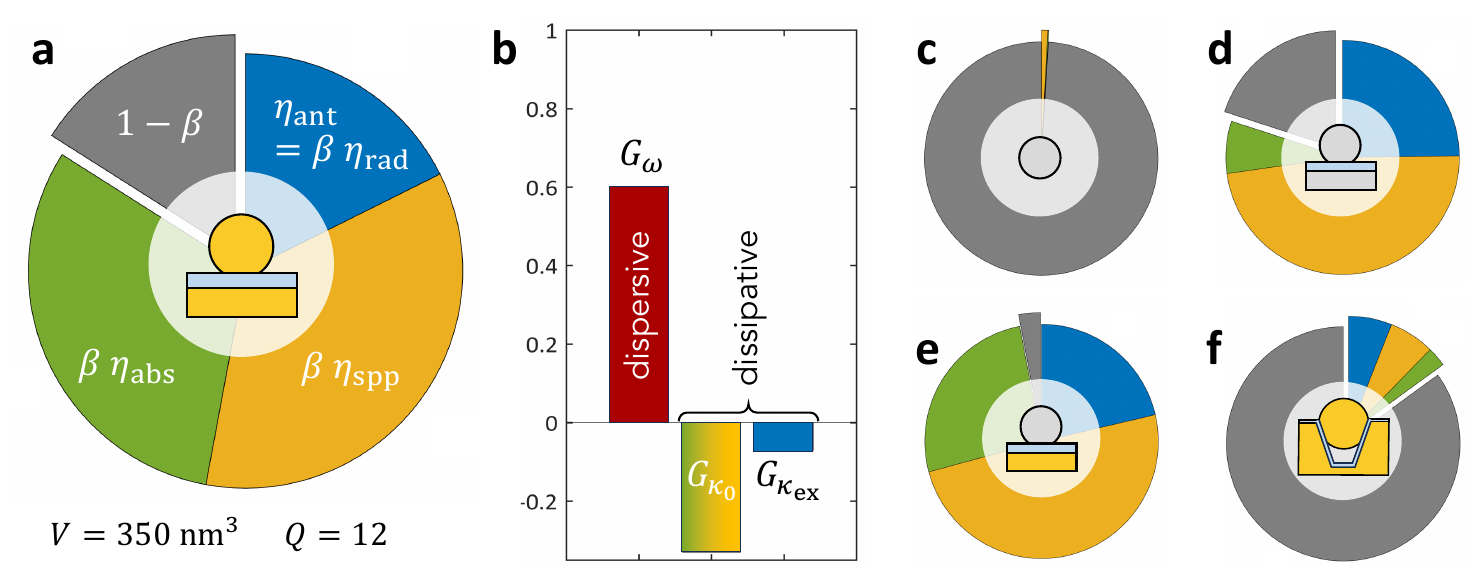}
    \caption{\footnotesize (a) Breakdown of decay channels for a dipole emitter optimally coupled to the bright mode of an NPoM cavity. The share $1-\beta$ corresponds to quenching, it is not a modal quantity (but depends on wavelength and position of the emitter). Mode volume is computed at the position of highest dipolar coupling to the mode, and $Q$ factor is evaluated by the spectral linewidth. Results were confirmed by QNM calculations. 
    (b) Relative contributions of dispersive and dissipative coupling rates, estimated by varying the permittivity in the nanogap and computing its effect on $\omega_p$, $\kappa_\text{ex}$ and $\kappa_0$. 
    (c-f) Breakdown of decay channels for the dominant bright mode of several nanostructures depicted as insets. A more complete list of parameters for various geometries is provided in Fig.~\ref{fig:listexample}, Appendix.
    }
    \label{fig:example}
\end{figure}
\subsection*{Nanocavity examples}

To practically illustrate the general formalism we present quantitative results in Fig.~\ref{fig:example} computed for a few typical geometries, starting with a gold nanoparticle-on-mirror (NPoM) cavity on a gold substrate where the Raman active molecules form a monolayer acting as a nanometric spacer. We stress here that some values, such as the mode volume and quenching rate, depend on the emitter position, which is here chosen so as to minimize mode volume. Accurately treating the collective optomechanical coupling of the Raman-active monolayer requires more advanced approaches (e.g., based on the Green's function as presented in Ref.~\citenum{jakob2023}) and remains a topic for future research. We remark in Fig.~\ref{fig:example}a that in spite of the nanometric distance between molecule and metal, the quenching of Raman emission is limited to $1-\beta\simeq 16\%$, an attractive property of metallic nanogaps that has been discussed in Refs.~\citenum{akselrod2014,faggiani2015,kongsuwan2018,zhang2019,baumberg2019} for example. The mode volume is 350~nm$^3$, close to the physical volume between the gold mirror and the facet of the nanoparticle. With a $Q$ factor of 12 at 760~nm, this dipolar bright mode decays into three channels: radiation (share $\eta_\text{rad}=\frac{\kappa_\text{ex}}{\kappa}$), internal losses by local absorption in the metal ($\eta_\text{abs}$) and losses as surface plasmon polarition (SPP) emission ($\eta_\text{spp}$), with $\eta_\text{abs}+\eta_\text{spp}=\frac{\kappa_0}{\kappa}$. The share $\eta_\text{spp}$ is here counted as a non-radiative loss channel, but proper engineering of the metallic substrate (e.g., with gratings \cite{long2016}) may convert SPP into free-space radiation, in which case $\eta_\text{rad}$ would increase accordingly, providing a substantial lever for improving radiative coupling efficiency in NPoMs. 
The input coupling efficiency depends on the excitation beam parameters, in particular its focusing and polarization; it is estimated as $\eta_\text{in}\sim 6\%$ only, which should be taken as an upper bound for an optimally polarised and focused beam \cite{vento2023a} and in the absence of excitation through SPPs.

Figure~\ref{fig:example}b summarizes the relative magnitudes of dispersive and dissipative coupling rates. Beware that they should not be summed up to obtain a total optomechanical coupling rate since they have distinct impacts on the system's dynamics. \cite{wu2014} These results were obtained by varying the permittivity $\varepsilon$ of the dielectric layer representing the molecules in the simulation and monitoring the effect on the resonance frequency and decay rates of the nanocavity mode of interest, with $G_\omega\propto \frac{\partial \omega_p}{\partial \varepsilon}$,  $G_{\kappa_0}\propto \frac{\partial \kappa_0}{\partial \varepsilon}$ and $G_{\kappa_\text{ex}}\propto \frac{\partial \kappa_\text{ex}}{\partial \varepsilon}$. This approach is simpler than the one employed in Ref.~\citenum{primo2020} where a single molecule was modelled by a dielectric ball of varying radius, but confirms that dissipative couplings are of similar magnitude as the dispersive coupling in McOM and should not be neglected (yet they have been in most works to date). 

We present other examples of nanocavity geometries in Figs.~\ref{fig:example}c-f. The single silver sphere antenna in Fig.~\ref{fig:example}c was simulated using both COMSOL and analytical Mie theory, yielding very similar results (see also Fig.~\ref{fig:listexample}). It is characterised by a bright dipolar mode having very good radiative efficiency $\eta_\text{rad}\simeq 85\%$, yet presents a high quenching rate $1-\beta=99\%$ (for a molecule 1~nm away from the sphere), illustrating the important distinction that should be made between the modal brightness $\eta_\text{rad}$ and what is often called the antenna efficiency $\eta_\text{ant}= \beta \eta_\text{rad}$ when quenching is a dominant loss channel (small $\beta$). \cite{wu2021} Two other flavours of NPoMs are presented in Figs.~\ref{fig:example}d,e to illustrate how the choice of metals may affect the shares of quenching and absorption. The last example in Fig.~\ref{fig:example}f is the nanoparticle-in-groove used in Ref.~\citenum{chen2021} as dual resonant cavity for mid-infrared upconversion. This design suffers from dominant quenching, probably related to the large size of the nanoparticle (150~nm), calling for improvements. The full list of simulated parameters for these geometries is provided in the Appendix, Fig.~\ref{fig:listexample}.

\subsection*{Pico- and nanocavities}

A special class of systems to which the theory of McOM was employed are so-called picocavities, which are believed to be atomic-scale hotspots randomly created in laser-illuminated nanocavities during SERS measurements \cite{benz2016}. Their phenomenology consists of a set of suddenly appearing new Raman peaks (the `picocavity signal') under continuous excitation, whose fully correlated fluctuations point to a single-molecule origin. They coexist with the usual SERS peaks (the `nanocavity signal') whose intensities generally remain weakly affected by the appearance of a picocavity. 
The observed linear increase of the anti-Stokes to Stokes intensity ratio of the picocavity peaks with laser power was consistent with the prediction of quantum backaction by the McOM theory \cite{benz2016}. (As mentioned earlier, McOM provides a set of rate equations for quantum backaction that are fully equivalent to those for vibrational pumping in SERS. \cite{kneipp1996}) This observation implies that 
the cooperativity $\mathcal{C}$ is comparable to the thermal occupancy of the vibrational mode ${n}_\text{th}$. 

Under the assumptions used in Ref.~\citenum{benz2016} that the decay rates $\kappa$ and $\gamma$ and the intracavity photon number $n_p$ are identical for nanocavity and picocavity, and with Raman polarizability estimated from DFT calculations of the bare molecule, the observation of vibrational pumping is consistent with an effective picocavity mode volume $V_\text{pico}$ on the order of 1~nm$^3$. Such values, which are two to three orders of magnitude smaller than those of nanocavities, have been also predicted by electromagnetic simulations \cite{urbieta2018,li2021,wu2021a,baumberg2022} (Fig.~\ref{fig:example}). 
We stress here that McOM is not a theory intended to explain the formation of picocavities.

If we now apply the scaling law presented above, which states that the cooperativity scales as $Q/V$ in McOM for filled cavities, we arrive at the intriguing prediction that the Raman signal from a picocavity could be two or three orders of magnitude stronger than the stable nanocavity signal. We emphasize here that this prediction already accounts for the fact that hundreds of molecules contribute to the nanocavity signal whereas just one does for the picocavity signal. Such giant Raman intensity enhancement has so far not been observed: experiments reported picocavity signals of similar magnitude as the nanocavity signal, which itself shows little modification during a picocavity event \cite{benz2016,shin2018,chen2021a,griffiths2021,griffiths2021a,baumberg2022,lin2022,poppe2023}. There could be multiple explanations for this apparent discrepancy: On the one hand, molecules in picocavities could possibly suffer from stronger quenching, reducing $\eta_\text{rad}$. On the other hand, a chemical enhancement mechanism could play a dominant role in explaining picocavity spectra, as suggested by DFT calculations that include just a few gold atoms around the molecule.\cite{lin2022} Finally, it is believed that large field gradients at the scale of the molecule are an essential feature of picocavities \cite{benz2016}, so that this type of coupling should be properly integrated in the McOM model, originally built around the dipole approximation.
In any case, it would be a worthwhile undertaking to gain better understanding of all parameters that enter the expression of detected Raman photon flux in such systems (Fig.~\ref{fig:etas}).


\subsection*{Experimental challenges}

Multiple of the physical phenomena predicted by the McOM theory await solid establishment in experimental demonstrations. To understand the challenges, we should first detail the main experimental requirements for rigorously applying the McOM formalism and test its predictions. A first set concerns the knowledge of the system: 

\begin{itemize}
    \item Single nanocavity measurements are to be privileged; if measuring an ensemble one must be certain that all nanocavities and their molecular contents are absolutely identical (which is difficult in practice given the nanometric critical dimensions involved). 
   
    \item The optical response of the nanocavity should be fully known in the frequency region covering the laser and Raman sideband. Experimentally, dark-field scattering spectroscopy provides valuable information at the single particle level, yet it is only sensitive to bright modes that efficiently scatter an incoming plane wave and does not suffice to reconstruct the near-field response that also depends on dark modes. It is also not able to distinguish the contributions from $\kappa_\text{ex}$ and $\kappa_\text{0}$ to the observed $Q$-factor. Numerical simulations can be helpful to obtain the decay rates $\kappa_\text{ex}$, $\kappa_0$, the mode volume $V$ and the bright mode branching ratio $\beta$. Simulations are quite accurate for gaps larger than 1~nm but require a precise knowledge of the geometry and refractive index of the molecular spacer (for gap nanocavities). Usually, these parameters are fine-tuned in the simulations to reproduce the measured dark-field scattering spectra. \cite{ahmed2021}  Electron beam methods offer nanoscale insights into the plasmonic modes, \cite{kociak2014} but NPoM geometries and vertical nanogaps in general are not readily accessible.  
    
    \item To obtain the coupling rate $g_0$ and single-photon cooperativity $\mathcal{C}_0$, knowledge of the Raman polarisability and the exact number of embedded molecules is required. In reality, each molecule has a different coupling rate depending on its exact location due to the inhomogeneity of the near-field, which can be treated numerically. \cite{jakob2023} But the main source of uncertainty here is the single-molecule Raman polarisability, and more particularly, how much it is affected by `chemical' enhancement for molecules bound to the metal. \cite{wu2008,chen2020,dao2021,leru2024} Accurate calculations of Raman cross-sections for large enough metal clusters coupled to small organic molecules remain difficult and they suggest that the chemical enhancement factor is usually large. \cite{lin2022} Alternatively, a more controllable thin material such as graphene \cite{lai2018} could be used as an intermediate intermediate layer between the plasmonic metal and the molecules, in an attempt to better separate chemical and electromagnetic enhancements.  
    
    \item Last, but not least, the intracavity  plasmon number $n_p$ must be estimated for a given pump power, which requires the additional knowledge of $\eta_\text{in}$.  Focused cylindrical vectorial beams allow optimising the power coupled to specific nanocavity modes, \cite{long2016,vento2023a} but a quantitative knowledge of  $\eta_\text{in}$ remains elusive. Obtaining this parameter from a simulation implies the use of a focused beam as background field, often much harder to implement than a plane wave excitation \cite{zhang2009}, in particular for nanocavities built on mirrors \cite{tang2021}. Experimental measurements of $\eta_\text{in}$ or $n_p$ are challenging in the context of McOM and SERS. Interferometric scattering microscopy \cite{taylor2019} was successfully employed to quantify the extinction of nanoantennas on glass \cite{gennaro2014,khadir2020} and should therefore allow retrieving $\eta_\text{in}$ in the context of McOM, where most cavities are assembled on a mirror substrate, though. Alternatively, and as elaborated below, the experimental geometry can be modified to resemble more that of the simulated plane wave excitation.
\end{itemize}

In addition to these challenges linked with the difficulties to fully characterise the near-field response of a gap nanocavity, another hurdle in the quest for McOM effects and applications is the limited tolerance of metallic nanocavities to large excitation powers. In Ref.\citenum{jakob2023}, for example, irreversible changes to the Raman spectra (which can be associated to optical `damage' in this context, Fig.~S23 in Ref.~\citenum{jakob2023}) represent a non-negligible portion of the studied power-dependent effect -- even though much care was taken to limit this damage by short laser exposure time and optimized pulse duration. 
The most exciting predictions of McOM are realised in the regime where $\mathcal{C}=n_p\mathcal{C}_0>1$, so that the intracavity plasmon number $n_p$ must reach the largest possible values -- at least be of order unity. To avoid excessive heating, this regime has been mostly explored under picosecond pulsed excitation. \cite{lombardi2018,crampton2018,liu2021,xu2022,jakob2023} 
This causes at least two types of difficulties. First, because $\eta_\text{in}$ is usually much smaller than one, most of the incoming laser power is not even coupling to the nanocavity mode of interest, but it can nevertheless be absorbed in the focal area and generate unwanted heat, hot-electrons, etc. 
Second, as $n_p$ increases inside the nanocavity, other nonlinearities may emerge and are hard to experimentally distinguish from potential optomechanical effects. In particular, both thermal effects \cite{ahmed2021,sun2022,lee2024} and electrons excited in the metal \cite{perner1997,hartland1999,wang2015,zhang2018} are expected to modify the transient plasmonic response on time scales down to picoseconds \cite{kumar2016,wang2017,schirato2022} -- comparable to the vibrational decay rate that dictates the optomechanical response. 

In fact, it remains an open question whether optomechanical nonlinearities can be observed before the onset of plasmonic nonlinearities. In most studies of ultrafast plasmonics on metallic nanoparticles, measurable nonlinear response (such as resonance shift and broadening) is observed for pump fluences as low as $100~\mu$J/cm$^2$ at off-resonant near-infrared wavelengths (see for example Refs.~\citenum{kumar2016,zhang2018,schirato2022}, among others).
When using a high pulse repetition rate of $\sim 100$~MHz typical of Ti:Sapph oscillators, this fluence translates in an average power of $100~\mu$W/$\mu$m$^2$. Considering a diffraction-limited spot-size in a confocal microscope with numerical aperture of 0.8-0.9, and that excitation is usually resonant with a plasmonic mode in McOM experiments, it means that few tens of microwatts -- or even less -- averaged power may be enough to drive a significant nonlinearity in the plasmonic response. In comparison, the highest average power used in Ref.~\citenum{jakob2023} was 60~$\mu$W. 
The optomechanical effects being investigated should therefore be carefully deconvoluted from the nonlinear plasmonic response itself, motivating further work. 

\subsection*{Theoretical challenges}

\begin{itemize}
    \item In contrast with macroscopic and mesoscopic mechanical resonators, molecular vibrations feature markedly anharmonic potentials, in particular for modes that are localized to few molecular bonds \cite{vento2023}. The standard McOM Hamiltonian, however, assumes a harmonic potential for the vibration. One recent theoretical work extends the formalism to account for vibrational anharmonicity in the case of a single, off-resonant molecule \cite{schmidt2024}. When sufficiently sharp features are present in the LDOS of the nanocavity, this work predicts a regime of incoherent mechanical blockade, provided that only the Stokes transition to the first excited vibration can be efficiently pumped by SERS, while the second one is not. Such sharply changing LDOS can be achieved for example by the Fano lineshape of a hybrid plasmonic-dielectric cavity \cite{doeleman2016,doeleman2020}. Another finding of Ref.\citenum{schmidt2024} is that anharmonicity has deep impacts on the regime of dynamical backaction amplification: it changes the threshold for coherent mechanical oscillations and lowers their amplitudes. A recent proposal also suggests that vibrational anharmonicity is a resource in single-molecule optomechanics for the production of antibunched photons through optomechanical blockade.\cite{moradikalarde2024} It is likely, however, that such effects rapidly disappear in the experimentally relevant setting of many molecules, since the collective modes should become harmonic.\cite{moradikalarde2024} Solving a model that includes anharmonicity and collective effects remains an open challenge. 
    We also mention here the case of strongly driven picocavities, where vibrational pumping combined with anharmonicity can lead to qualitatively similar Raman peak shifts as predicted from the optical spring calculation. \cite{jakob2023}
    
    \item Primo \textit{et al.} put forward that dissipative optomechanical coupling can be on par with dispersive coupling in McOM.\cite{primo2020} Dissipative coupling occurs when the mechanical displacement modifies one of the dissipation rates of the cavity ($\kappa_\text{ex}$ and/or $\kappa_0$) \cite{weiss2013,wu2014}. The formula for $g_0$ originally derived in Refs.~\citenum{roelli2016,schmidt2016a} does not capture dissipative coupling, which is intimately related with the multimode and non-hermitian nature of nanocavities.\cite{yanay2016}
    The presence of dissipative coupling impacts the shape and magnitude of the optomechanical damping rate vs. laser wavelength \cite{weiss2013,wu2014}.  Ref.~\citenum{primo2020} applies perturbation theory in the quasi-normal mode formalism to compute an imaginary part of the optomechanical coupling rate that corresponds to dissipative coupling, confirming the results with full-wave moving-mesh simulations. They conclude that dissipative coupling is at least of comparable magnitude as dispersive coupling in the canonical NPoM nanocavity, which we confirm using a simplified approach in Fig.~\ref{fig:example}b.  It is also expected to prominently feature in engineered multimode McOM systems, like hybrid nanoantenna-cavity resonators.\cite{shlesinger2023}
    
    A point that has not been explicitly clarified in the literature to date is whether the Green's function approach to McOM \cite{kamandardezfouli2017,zhang2021}, that can treat arbitrary plasmonic and dielectric resonators, implicitly accounts for dissipative coupling. The difficulty here is that the Green's function approach does not yield an optomechanical coupling rate $g_0$ but directly provides instead expressions for the dynamics of vibrational and plasmonic operators.   
    
    \item Resonant and near-resonant Raman scattering could in first approximation be thought of as just increasing the Raman polarizability that enters the McOM model desribed in the first part. But recent theoretical work by M. Martinez-Garcia \textit{et al.} concluded that, even in the weak optomechanical coupling regime, interference between resonant and nonresonant contributions to SERS can deeply alter the predicted nanocavity scattering spectrum.\cite{martinez-garcia2024} Previously, S. Hughes \textit{et al.} already investigated the quantum nonlinear regime of resonant strong coupling between the nanocavity mode and the molecular electronic transition, in the presence of vibrational modes \cite{hughes2021}. Future experimental and theoretical works are needed clarify the precise impact of  molecular electronic resonance on vibrational dynamics and nanocavity scattering spectrum in the context of accessible SERS scenarios. 
    
    \item As already discussed above, experimentally determining the intracavity plasmon number remains an open challenge; conversely, numerical estimates are usually performed under the simplifying assumption of plane wave excitation and there is little quantitative work that accounts for strongly focused laser beams, possibly with complex vectorial fields \cite{vento2023a}. Such simulations do exist \cite{zhang2009,tang2021} but have not been implemented yet in the context of McOM to obtain more accurate estimates of input coupling efficiency and intracavity plasmon number. 
    
    \item A resource-efficient formalism that correctly accounts for collective effects among molecules coupled to a same nanocavity is still missing. The pioneering calculations presented in Ref.~\citenum{jakob2023} require numerical evaluation using a discrete set of point dipoles representing the molecules, limiting this approach to 100 molecules so far, while typical nanocavities can host 1000 or more. Since 2D materials can also be used as Raman active layers in McOM \cite{liu2021,xu2022}, the development of a continuum phonon model that applies also for molecular layers and accounts for collective dynamics would be valuable. Additionally, it was recently proposed that dipole-dipole coupling among closely packed IR-active molecules is an important source of collective behavior \cite{chen2019,gray2021} having impacts on the Raman spectra \cite{mueller2022}; it would be appealing to include such effects in a comprehensive McOM description. 
    
\end{itemize}


\subsection*{Molecular optomechanically induced transparency}


\begin{figure}
    \centering
    \includegraphics[keepaspectratio, width=0.9\textwidth]{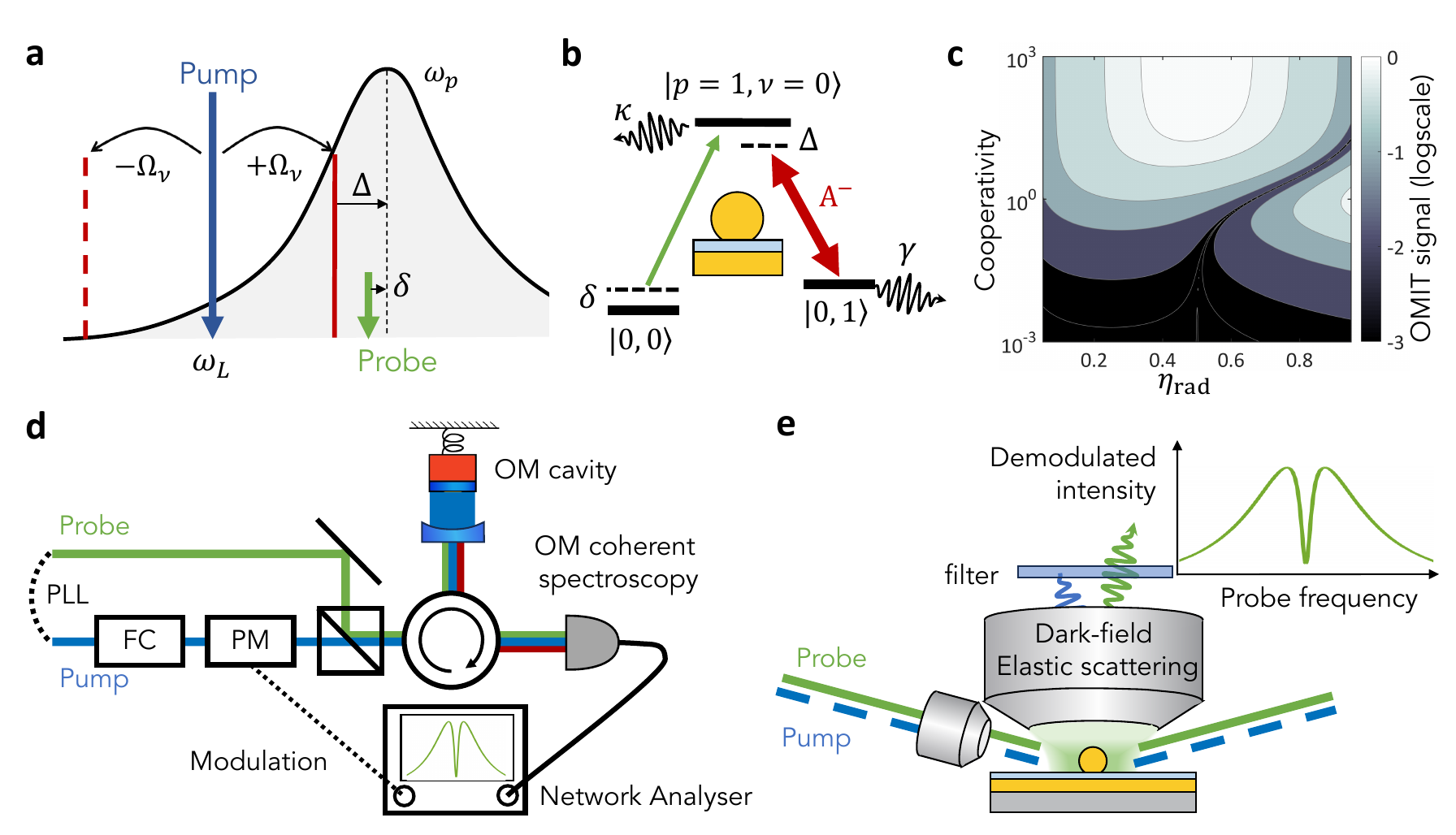}
    \caption{\footnotesize Optomechanically induced transparency: concept and measurement schemes. (a) Frequency representation of all involved fields and resonances, with the definitions of detunings $\Delta$ and $\delta$. (b) Same parameters represented on an energy level diagram in the single excitation sub-space. The notation $|p,\nu\rangle$ designates a state with $p$ plasmons and $\nu$ vibrations. (c) Computed map of OMIT signal strength (normalised to the input probe power) vs. cooperativity 
    $\mathcal{C}$ and external coupling efficiency $\eta_\text{rad}$. (d) Possible OMIT measurement scheme in macroscopic cOM, adapted from Ref.~\citenum{qiu2020}. PLL: phase-locked loop; FC: filter cavity; PM: phase modulator. (e) Proposed implementation of OMIT measurement in McOM using a dark-field geometry for pump and probe excitation. A spectral filter blocks the pump and the probe signal is sent to the detector. The pump (pulsed or cw) is intensity-modulated so that the demodulated scattered probe intensity senses the pump-induced change in cavity scattering, which can have thermal (broad plasmonic response) or optomechanical (narrow vibrational response) origins.  
    }
    \label{fig:OMIT}
\end{figure}


Contrary to macroscopic cavity optomechanics, where heterodyne linear optical field measurements are the norm, McOM has been so far only studied with photon-counting detectors such as CCD cameras, possibly losing valuable phase information in the process. In particular, depending on the detuning between laser and cavity resonance the optomechanical (Raman) sidebands are predicted to consist of different contents of phase and amplitude modulations: Pure phase modulation is expected on resonance, which can be readout by homodyne or heterodyne detection. We therefore suggest to introduce the use of local oscillators generated by a nonlinear optical process in the measurement of SERS signals from nanocavities, as already implemented in the case of stimulated Raman scattering (SRS) imaging \cite{potma2006} and vibrational sum frequency generation \cite{stiopkin2008}. 

Related, a valuable measurement that was not yet implemented as such in McOM is that of optomechanically induced transparency, or OMIT \cite{weis2010,safavi-naeini2011}. Beyond its fundamental appeal and potential applications \cite{xiong2018}, OMIT measurements allow a very direct estimate of the cooperativity, a capability that is still lacking in McOM. 
The effect is described in Fig.~\ref{fig:OMIT}a,b: a strong pump laser is tuned on the lower vibrational sideband of the cavity, so that the anti-Stokes frequency is near resonant with the cavity frequency (their detuning is denoted as $\Delta$). A weaker probe laser beam is swept in frequency (detuning $\delta$ from the cavity). In macroscopic cOM, a destructive interference between the two excitation pathways depicted in Fig.~\ref{fig:OMIT}b can induce a transparency window for the probe when $\Delta\simeq \delta$ compared to its absorption by the cavity in the absence of pump field. A detailed calculation adapted to the specificity of McOM (where scattered light is detected instead of reflection or transmission) has not been published to our knowledge; however we expect that a sharp feature is expected in the probe scattering spectrum (peak or dip) at $\Delta\simeq \delta$. Panel (c) in Fig.~\ref{fig:OMIT} estimates the relative strength of this feature (for the case $\Delta=0$) as a function of cooperativity $\mathcal{C}={n}_p \mathcal{C}_0$ (where ${n}_p$ scales linearly with pump power) and external coupling efficiency $\kappa_\text{ex}/\kappa=\eta_\text{rad}$. Two regimes are favorable to the observation of OMIT: (i) overcoupled cavity ($\eta_\text{rad}\sim 1$) with moderate cooperativity ($\mathcal{C}\simeq 1$), or more conventionally (ii) high  cooperativity ($\mathcal{C}\gg1$) under critical coupling ($\eta_\text{rad}= 0.5$).

A typical OMIT measurement scheme in macroscopic cOM is presented in Fig.~\ref{fig:OMIT}d; how should OMIT be measured in the context of McOM on a plasmonic nanocavity? The first challenge is that plasmonic nanocavities are not probed in reflection or transmission given their subwavelength dimensions, but rather in scattering geometry. What matters is to isolate the optical field that has interacted with the nanocavity mode. For that, we propose a dark-field scattering detection geometry (Fig.~\ref{fig:OMIT}e) where the specular reflection of pump and probe beam are not collected. The pump frequency is filtered out of scattered field and the probe power is monitored; modulating the pump intensity and demodulating the probe signal using lock-in detection should allow to reach the required sensitivity for observing OMIT with state-of-the-art nanocavities (Fig.~\ref{fig:OMIT}c). Compared to recent implementations of SRS on plasmonic antennas \cite{frontiera2011,prince2017}, the proposed scheme mainly differs in the dark-field detection geometry and the prerequisite to be able to observe the absorption or scattering of a single nanocavity; also, note that the pump and probe beams in OMIT correspond to the Stokes and pump beams in SRS, respectively.  This measurement could be performed under ps pulsed or continuous wave excitation.

\section*{Conclusion}

By reformulating the physics of SERS in the language of cavity optomechanics, the framework of molecular cavity optomechanics opens a new arena for understanding fundamental aspects of plasmon-enhanced Raman scattering and for applying this phenomenon to nonlinear nano-optics and possibly quantum technologies. In this Perspective, we first clarified the content of this framework and how to apply it to typical nanocavities used for SERS. We identified the need for a better characterisation of the input coupling efficiency and different dissipation channels of the nanocavity field, and recalled that the strength of optomechanical effects should be quantified by the cooperativity instead of the coupling rate alone. The methodology for calculating these quantities was illustrated on a couple of concrete and relevant examples, and a partial overview of nanocavity performance compared to dielectric microcavities was proposed. A number of experimental challenges were listed, which still prevent unambiguous identification of and control over predicted optomechanical effects in SERS such as dynamical backaction amplification, optical spring, or collective vibrational response of molecular ensembles. Some fundamental questions remain open at the level of the theory itself as well, such as how to include vibrational anharmonicity, how to properly account for dissipative coupling, or how to formulate a continuum model of the vibrational mode for many molecules or 2D materials. 


Looking ahead, we can see numerous promising and challenging research directions. Hybrid plasmonic-dielectric approaches may allow to reach overcoupled cavities as well as deeply sideband-resolved McOM. New ideas are needed to prevent metal surface restructuring under high laser power, and the failure mode under pulsed excitation should be clarified. 
Molecules with long-lived vibrational modes and large Raman cross-section should be engineered to boost the cooperativity. Entanglement between plasmon and molecular vibrations, and among vibrating molecules, should be demonstrable. Exploiting resonant or near-resonant coupling to electronic molecular transitions would establish tripartite phonon-photon-exciton systems \cite{reitz2019,wang2019,zhao2020,neuman2020,hughes2021,nation2023}, bearing analogy to macroscopic optomechanical systems that include two-level systems, with potential for enhanced coupling strengths, nonlinear effects, and applications in wavelength conversion \cite{whaley-mayda2021,chikkaraddy2023a}. Finally, the realization of the single-photon optomechanical strong coupling would give rise to a nonlinear response at the quantum level, both for photonic and phononic degrees of freedom, a holly grail in quantum optomechanics \cite{rabl2011,nunnenkamp2011}.

\begin{acknowledgement}

We thank Femius Koenderink and Ilan Shlesinger for fruitful discussions and critical reading of the manuscript. 

\end{acknowledgement}

\section*{Funding Sources}
This work is part of the Research Program of the Netherlands Organisation for Scientific Research (NWO). This work has received funding from the European Union's Horizon 2020 research and innovation program under Grant Agreements No. 829067 (FET Open THOR) No. 820196 (ERC CoG QTONE). 
P. R. acknowledges financial support from the Swiss National Science Foundation (Grant No. 206926)
and from the European Union's Horizon 2020 research and innovation programme under the Marie Skłodowska-Curie grant agreement No 101065661. 
C.G. acknowledges the support from the Swiss National Science Foundation (project numbers 214993 and 198898).




\newpage
\section{Appendix}


\begin{figure}
    \centering
    \includegraphics[keepaspectratio, width=1\textwidth]{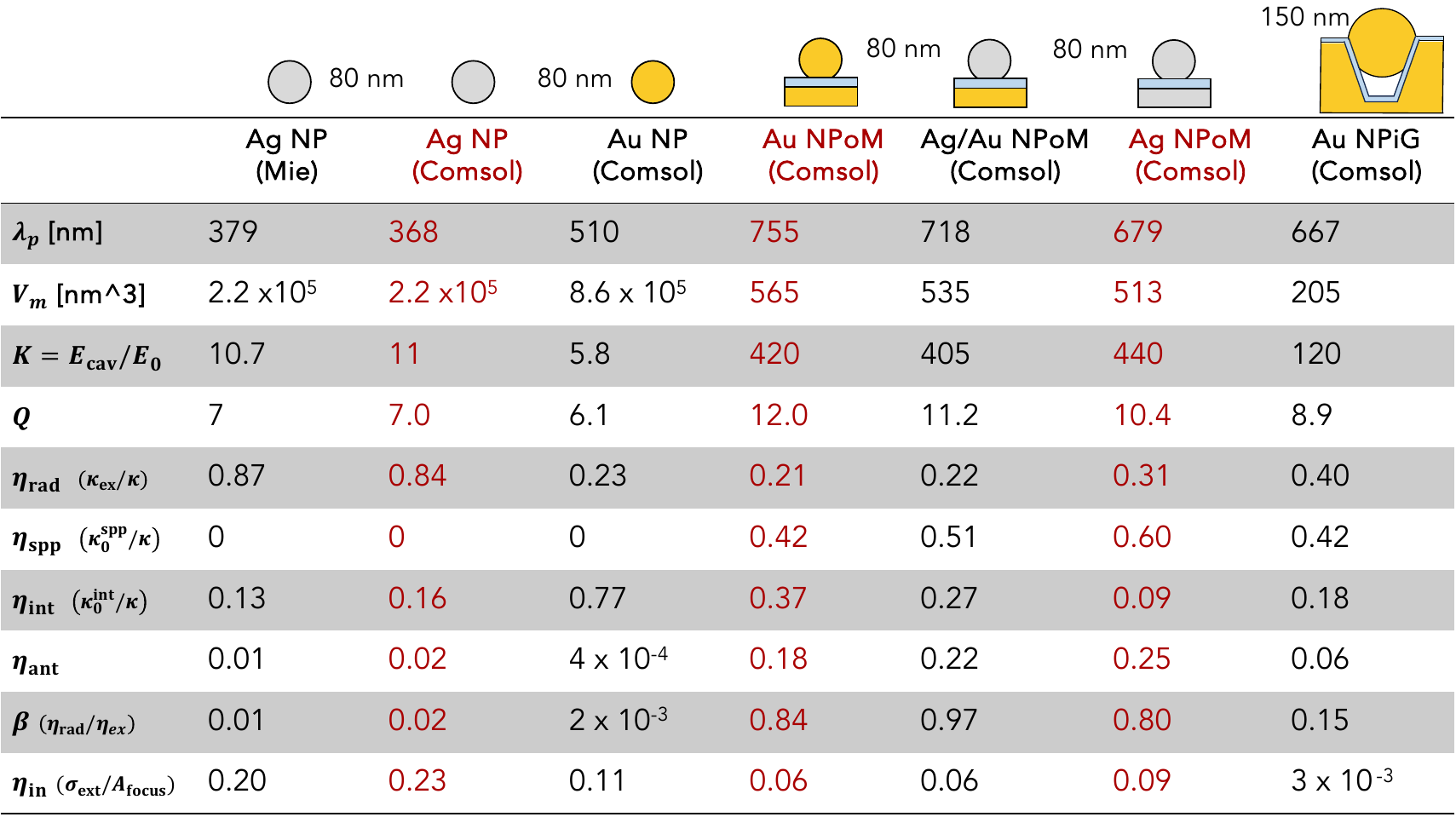}
    \caption{\footnotesize Full list of parameters extracted from the simulations of various plasmonic nanostructures. 
    The parameters characterize the dipolar plasmonic mode for the single NP cases, 
    the L01 mode for the NPoM cases and the anti-symmetric L01 mode for the NP in groove (NPiG) case. 
    $E_{\mathrm{cav}}$ is here defined as the spatial maximum of the norm of the electric field at a distance $d=1$~nm from the nanoparticle's surface 
    and $E_\mathrm{0}$ is the norm of the background electric field generated or by a plane-wave excitation or by a point-dipole source. For nanogap structures (NPoM and NPiG), $E_{\mathrm{cav}}$ is taken on the mid-gap plane, which is at a distance of 0.65~nm from the surface considering a single layer of molecules 1.3~nm thick.
    To calculate $A_{\mathrm{focus}}$, a focusing objective of numerical aperture $\mathrm{NA}=0.9$ is considered. 
    The NPs are modelled by a sphere of 80~nm diameter. 
    The NPoMs are composed by a truncated sphere of 80~nm with a facet size of 10~nm, a 300~nm thick metallic film separated from the sphere by a dielectric layer (1.3~nm thickness, refractive index of 1.4). 
    For the NPiG case, the dimensions of the truncated sphere are modified (150~nm diameter without facets) and the groove dimensions considered are as follows: 1.4 $\mu$m long with a trapezoidal cross-section whose bases are 180~nm and 80~nm, the height is 150~nm. The corners were rounded by 20~nm.
    For plane-wave excitation simulations of the nanogap structures, we consider a incident wave at an angle of 60 deg from the gold plane's normal (p-polarized for the NPoM case, orthogonal to the groove for the NPiG case). 
    }
    \label{fig:listexample}
\end{figure}


This appendix provides additional guidelines regarding the calculation of the cavity parameters introduced in this Perspective. 
It has to be noted that many aspects regarding calculations of cavity parameters for plasmonic cavities have been covered in depth elsewhere 
\cite{faggiani2015,roelli2016,schmidt2016a,wu2021,leru2024}. 
Below, we focus on a simple case study (a small spherical nanoparticle, NP) for which a Mie theory approach is possible. 
The spherical NP case, being extensively investigated in the literature, constitutes the most elemental way to introduce cavity parameters for the plasmonics community. 
It also offers a natural benchmark to evaluate and connect the different frameworks existing in the literature. 
Building on the NP case, we conclude the Appendix with a presentation of the FEM simulations used to address more complex structures, such as the NPoM and NP-in-groove (NPiG) geometries. Some representative results are summarized in Fig.~\ref{fig:listexample}. 




\subsection*{A Mie approach: spherical nanoparticle}
\begin{figure}
    \centering
    \vspace{-12pt}
    \includegraphics[keepaspectratio, width=1\textwidth]{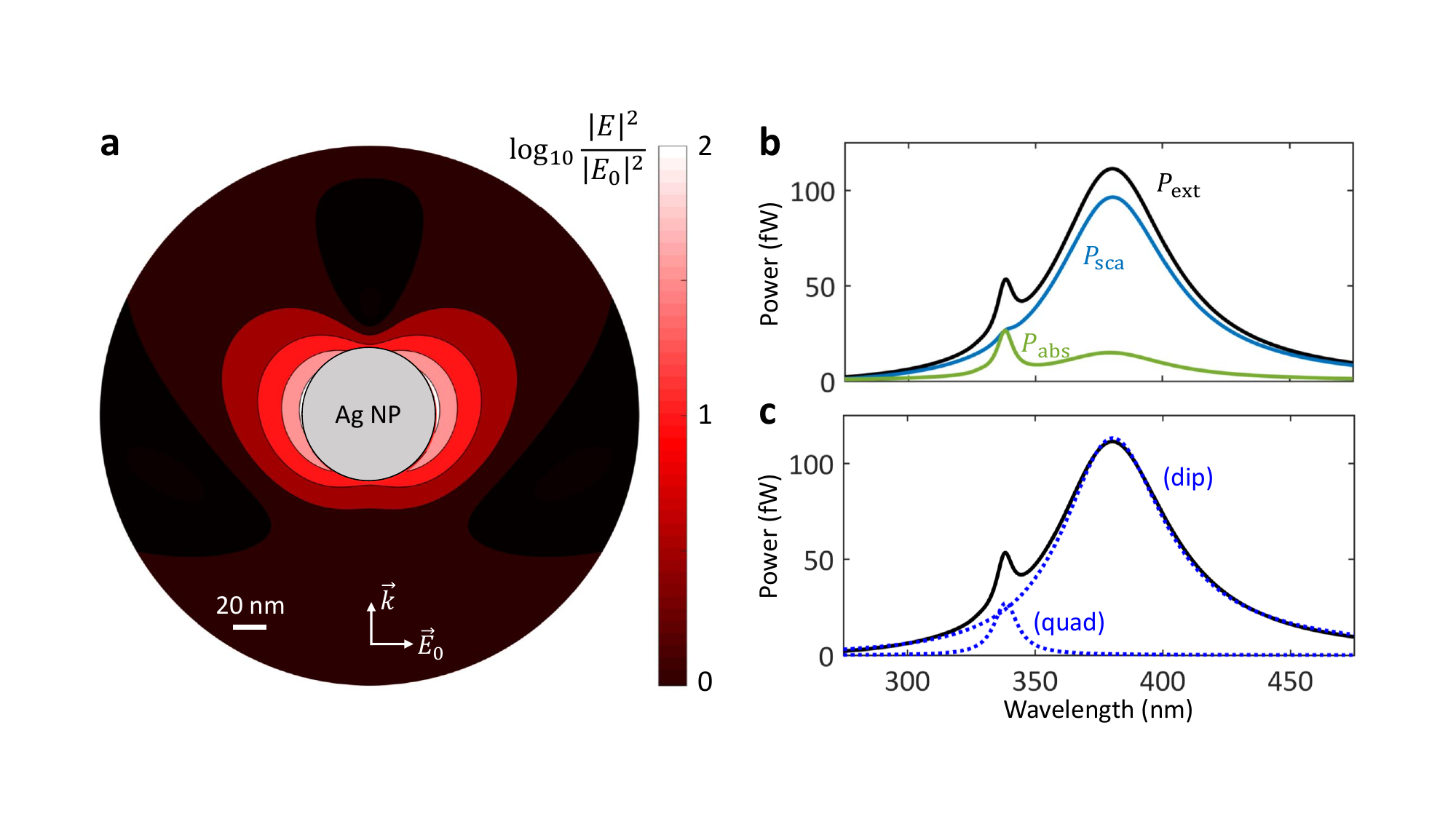}
    \vspace{-48pt}
    \caption{\footnotesize Mie calculations for a spherical silver NP of 80~nm diameter under plane-wave excitation. 
    (a) Spatial distribution of the intensity enhancement factor 
  around the silver NP for a plane-wave excitation tuned to the dipole resonance.
    (b) Spectra of extinguished, scattered and absorbed power. 
    (c) Multi-Lorentzian fit of the extinguished power, with the dipolar and quadrupolar contributions labelled (dip) and (quad), respectively. 
    }
    \label{fig:miesolver}
\end{figure}

The Mie theory gives an exact solution to the problem of the scattering of an electromagnetic (EM) wave by a sphere. 
In addition to the EM field solutions at any point in space, this theory conveniently provides analytical expressions 
for integral quantities such as the scattered power $P_{\mathrm{sca}}$ and the absorbed power $P_{\mathrm{abs}}$ (see Fig.~\ref{fig:miesolver}). 
For a sufficiently small silver nanosphere, all parameters of the plasmonic dipole mode can be accurately described using scattering expressions.  
Using a Mie solver package \cite{ru2008}, the spectra $P_{\mathrm{sca}}(\lambda)$, $P_{\mathrm{abs}}(\lambda)$ and $P_{\mathrm{ext}}(\lambda)= P_{\mathrm{sca}}(\lambda)+P_{\mathrm{abs}}(\lambda)$
are obtained by scanning the wavelength of the plane-wave excitation (PWE) 
at constant incident intensity ($S_{\mathrm{inc}}=P_{\mathrm{inc}}/A_{\mathrm{focus}}$, 
with $A_{\mathrm{focus}}$ the effective focal area of the incoming light away from the nanoparticle). 
Analytical expression of $A_{\mathrm{focus}}$ for different light waves and its connection with 
the normalized energy density of the wave, can be found in Ref.~\citenum{zumofen2008}. 


A multi-Lorentzian fit of $P_{\mathrm{ext}}(\lambda)$ determines the dipole resonance wavelength ($\lambda_{\mathrm{p}}$) and its decay rate ($\kappa$). 
Other single-mode cavity parameters are subsequently derived from direct evaluation of the Mie calculation results on resonance: 
the effective cross-section $\sigma_{\mathrm{ext}}(\lambda_{\mathrm{p}})=P_{\mathrm{ext}}(\lambda_{\mathrm{p}})/S_{\mathrm{inc}}$, 
the radiative efficiency $\eta_{\mathrm{rad}}\approx P_\mathrm{sca}(\lambda_{\mathrm{p}})/P_\mathrm{ext}(\lambda_{\mathrm{p}})$, 
the non-radiative efficiency $\eta_{\mathrm{abs}}\approx P_\mathrm{abs}(\lambda_{\mathrm{p}})/P_\mathrm{ext}(\lambda_{\mathrm{p}})$ 
and the local field enhancement factor $K(d)$ at a distance $d$ from the nanoparticle's surface. 
Implicitly, by applying a multi-Lorentzian fit (Fig.~\ref{fig:miesolver}c), we assume here that the multimode $P_{\mathrm{sca/abs}}$ can be decomposed into its single-mode components.  
To quantify the impact of other plasmonic modes on a point-like object at a distance $d$ from the NP's surface, we make use of Mie calculations with a point-dipole excitation (PDE). 
The (total and radiative) Purcell factors extracted from the PDE calculations, $M_{\mathrm{tot}}$ and $M_{\mathrm{rad}}$ respectively, enable to evaluate the antenna efficiency: 
$\eta_{\mathrm{ant}}=M_{\mathrm{rad}}/M_{\mathrm{tot}}$. 
Since all plasmonic modes contribute to $M_{\mathrm{tot}}$, the ratio $\beta=\eta_{\mathrm{ant}}/\eta_{\mathrm{rad}}$ 
quantifies the probability for the dipole to emit into the bright mode; the probability of emission into other modes, i.e. quenching is then given by $1-\beta$.

The remaining single-mode cavity parameters can be calculated from the following expressions.
\begin{itemize}
    \item  The mode volume $V\simeq \sigma_{\mathrm{ext}}c/(K^2\kappa)$ 
\cite{roelli2016}, with $c$ the speed of light in vacuum. 
    \item The mean intra-cavity photon number $n_p$ under laser illumination of power $P_{\mathrm{inc}}$ at frequency $\omega_L/(2\pi)$ detuned from the cavity resonance by $\Delta=\omega_L-\omega_p=\omega_L-2\pi c /\lambda_p$ \cite{aspelmeyer2014}:
$$n_p=\frac{\eta_{\mathrm{in}}P_{\mathrm{inc}}}{\hbar\omega_L}\frac{\kappa/2}{\Delta^2+\kappa^2/4}$$
with $\eta_{\mathrm{in}}=\sigma_{\mathrm{ext}}/A_{\mathrm{focus}}$ the input coupling efficiency. 
\end{itemize}
This simple approach can be applied to more elaborated structures via FEM and FDTD simulations as long as the targeted mode is sufficiently well-isolated (spectrally or spatially from the contributions of other modes). 

\subsection*{Estimation of the mode volume $V$: alternative method}
 The total Purcell factor $M_{\mathrm{tot}}$ calculated from the PDE calculations is dominated by a continuum of modes \cite{delga2014} even at the dipolar mode's resonance frequency so that direct mode volume calculation would not be possible.  
 On the contrary, for a PDE tuned in resonance with a sufficiently isolated cavity mode, 
 the radiative part of the Purcell factor $M_{\mathrm{rad}}$ is, to a very large extent, due to this mode only. 
 With the help of the efficiencies calculated previously, we can approximate the total Purcell factor due to a single dipolar bright mode as: 
$M^{\mathrm{(dip)}}_{\mathrm{tot}}=M_{\mathrm{rad}}(1+\eta_{\mathrm{abs}}/\eta_{\mathrm{rad}})=M_{\mathrm{rad}}(1+\kappa_{\mathrm{0}}/\kappa_{\mathrm{rad}})$.
This factor can be used directly to approximate the mode volume following the approach of Ref.~\citenum{schmidt2016a} or be linked to the approach exposed above via the following relation between Purcell and field enhancement factors: $K^2=\eta_{\mathrm{rad}}M^{\mathrm{(dip)}}_{\mathrm{tot}}$. 
It has to be noted that both mode volume estimates given here are in very good agreements with the values found by F. Koenderink in Ref.~\citenum{koenderink2010} for small spherical Ag NP and with our own evaluations via FEM calculations of the mode volume following a quasi-normal mode (QNM) treatment \cite{wu2021}. 
A similar level of agreement is found for the other cavity parameters calculated for the single NP case. 

\subsection*{Beyond the Mie theory approach}
A rigorous modal treatment, based on the quasi-normal mode formalism, is possible 
for many of the parameters introduced above \cite{wu2021}. 
The direct evaluation of $\eta_{\mathrm{rad}}$, named modal brightness $\mathcal{B}$ in the QNM framework, remains however a tedious task. 
Indirect calculations of $\eta_{\mathrm{rad}}$ for NPoM cavities have been introduced in Ref.~\citenum{faggiani2015}. Performing PDE calculations followed by near-to-far field transformation (NFFT) \cite{yang2016}, the energy dissipated through radiative and guided-mode (SPP) channels can be separated. This way, $\eta_{\mathrm{ant}}$ and $\beta\eta_{\mathrm{spp}}$ are found. Additional QNM calculations enable to evaluate $\beta$ so that the multimode absorption channel $\beta\eta_{\mathrm{abs}}=\beta-\eta_{\mathrm{ant}}-\beta\eta_{\mathrm{spp}}$ as well as all single-mode counterparts \{$\eta_{\mathrm{rad}},\eta_{\mathrm{spp}},\eta_{\mathrm{abs}}$\} can be quantified. 

In this Perspective, and in analogy with the single NP treatment discussed above, we perform PWE calculations on resonance with the plasmonic mode of interest followed by NFFT to evaluate the single-mode loss channels $\eta_{\mathrm{rad}}$ and $\eta_{\mathrm{spp}}$. Similarly to the treatment of Ref.~\citenum{koenderink2010}, we evaluate the divergence of the Ohmic losses with the size of the integration domain \cite{jiang2020} to separate the SPP contribution to absorption from the single-mode localized absorption ($\eta_{\mathrm{abs}}$). 

\bibliography{references_manual}
\end{document}


\title{Appendix - ACS Perspective on mOM}
\author{Philippe Roelli}
\affiliation{%
  CIC nanoGUNE, Nano-optics Group, E-20018 Donostia-San Sebastián, Spain
}%
\date{\today}

\maketitle



This appendix provides additional guidelines regarding the calculation of the cavity parameters introduced in this Perspective. 
It has to be noted that many aspects regarding calculations of cavity parameters for plasmonic cavities have been covered in depth elsewhere 
\cite{faggiani_quenching_2015,roelli_molecular_2016,schmidt_quantum_2016,wu_nanoscale_2021,le_ru_enhancement_2024}. 
Here, we focus on a simple study case (a small nanoparticle, NP) for which a Mie theory approach is possible. 
The NP case, being extensively investigated in the literature, constitutes the most elemental way to introduce cavity parameters for the plasmonics community. 
It also offers a natural benchmark to evaluate and connect the different frameworks existing in the literature. 
Extending on the NP case, we complete the Appendix with a presentation of the FEM simulations we used in order to tackle more complex structures, like the NPoM and NP-in-groove geometries. 




\subsection*{A Mie approach: single NP case}
The Mie theory gives an exact solution to the problem of the scattering of an electromagnetic (EM) wave by a sphere. 
In addition to the EM field solutions at any point in space, this theory conveniently provides analytical expressions 
for integral quantities such as the scattered power $P_{\mathrm{sca}}$ and the absorbed power $P_{\mathrm{abs}}$. 
For a sufficiently small silver nanosphere, all parameters of the plasmonic dipole mode can be accurately described using scattering expressions.  
Using a Mie solver package \cite{le2009sers}, the spectra $P_{\mathrm{sca}}(\lambda)$, $P_{\mathrm{abs}}(\lambda)$ and $P_{\mathrm{ext}}(\lambda)= P_{\mathrm{sca}}(\lambda)+P_{\mathrm{abs}}(\lambda)$
are obtained by scanning the wavelength of the plane-wave excitation (PWE) 
at constant incident intensity ($S_{\mathrm{inc}}=P_{\mathrm{inc}}/A_{\mathrm{focus}}$, 
with $A_{\mathrm{focus}}$ the effective focal area of the incoming light away from the nanoparticle). 
Analytical expression of $A_{\mathrm{focus}}$ for different light waves and its connection with 
the normalized energy density of the wave, can be found in Ref.~\cite{zumofen_perfect_2008}. 

A multi-Lorentzian fit provides the dipole resonance wavelength ($\lambda_{\mathrm{p}}$) and its decay rate ($\kappa$). 
Many single-mode cavity parameters are then obtained directly from the evaluation of the scattering parameters on resonance: 
the effective cross-section $\sigma_{\mathrm{ext}}(\lambda_{\mathrm{p}})=P_{\mathrm{ext}}(\lambda_{\mathrm{p}})/S_{\mathrm{inc}}$, 
the radiative efficiency $\eta_{\mathrm{rad}}\approx P_\mathrm{sca}(\lambda_{\mathrm{p}})/P_\mathrm{ext}(\lambda_{\mathrm{p}})$, 
the non-radiative efficiency $\eta_{\mathrm{0}}\approx P_\mathrm{abs}(\lambda_{\mathrm{p}})/P_\mathrm{ext}(\lambda_{\mathrm{p}})$ 
and the local field enhancement factor $K(d)$ at a distance $d$ from the nanoparticle's border. 

We assume here that under plane-wave excitation, 
the multimode $P_{\mathrm{sca/abs}}$ can be decomposed into its single mode components (here, we consider the dipole mode).  
To quantify the impact of other modes on an emitter/scatterer at a distance $d$ from the NP's border, we make use of Mie calculations with a point-dipole excitation (PDE). 
The (total and radiative) Purcell factors, $M_{\mathrm{tot}}$ and $M_{\mathrm{rad}}$ respectively, extracted from the PDE calculations define the antenna efficiency: 
$\eta_{\mathrm{ant}}=M_{\mathrm{rad}}/M_{\mathrm{tot}}$. 
Since all (dark) modes contribute to $M_{\mathrm{tot}}$, the ratio $\beta=\eta_{\mathrm{ant}}/\eta_{\mathrm{rad}}$ 
quantifies the probability for the dipole emitter to feed into the bright mode; in other words the quenching probability is $1-\beta$.

The remaining single-mode cavity parameters can be calculated from the following expressions.
\begin{itemize}
    \item  The mode volume $V\simeq \sigma_{\mathrm{ext}}c/(K^2\kappa)$ 
\cite{roelli_molecular_2016}, with $c$ the speed of light in vacuum. 
    \item The mean intra-cavity photon number $n_p$ under laser illumination of power $P_{\mathrm{inc}}$ and frequency $\omega_L/(2\pi)$ detuned from the cavity resonance by a factor $\Delta=\omega_L-\omega_p$:
$$n_{\mathrm{cav}}=\frac{\eta_{\mathrm{in}}P_{\mathrm{inc}}}{\hbar\omega_L}\frac{\kappa/2}{\Delta^2+\kappa^2/4}$$
\cite{aspelmeyer_cavity_2014}, with $\eta_{\mathrm{in}}=\sigma_{\mathrm{ext}}/A_{\mathrm{focus}}$ the input coupling efficiency.  

\end{itemize}
This simple approach can be applied to more elaborated structures via FEM and FTDT simulations as long as the targeted mode is sufficiently well-isolated (spectrally or spatially from the contributions of other modes). 

\subsection*{Estimation of the mode volume $V$: alternative method}
 The total Purcell factor $M_{\mathrm{tot}}$ calculated from the PDE calculations is dominated by a continuum of modes \cite{delga_quantum_2014} even at the cavity mode's resonance frequency so that direct mode volume calculation would not be possible.  
 On the contrary, for a point-dipole excitation tuned in resonance with a sufficiently well-isolated cavity mode, 
 the radiative part of the Purcell factor $M_{\mathrm{rad}}$ is, to a very large extent, due to this mode only. 
 With the help of the efficiencies calculated previously, we can approximate the total Purcell factor due to a single dipolar bright mode as: 
$M^{\mathrm{(dip)}}_{\mathrm{tot}}=M_{\mathrm{rad}}(1+\eta_{\mathrm{0}}/\eta_{\mathrm{rad}})=M_{\mathrm{rad}}(1+\kappa_{\mathrm{0}}/\kappa_{\mathrm{rad}})$.

This factor can be used directly to approximate the mode volume following the approach of Ref.~\citenum{schmidt_quantum_2016} or be linked to the approach exposed above via the following relation between Purcell and field enhancement factors: $K^2=\eta_{\mathrm{rad}}M^{\mathrm{(dip)}}_{\mathrm{tot}}$. 

It has to be noted that both mode volume estimates given here are in very good agreements with the values found by Koenderink in Ref.~\citenum{koenderink_use_2010} for small spherical Ag NP and with our own evaluations via FEM calculations of the mode volume following a quasi-normal mode (QNM) treatment \cite{wu_nanoscale_2021}. 
Same agreement is found for the other cavity parameters calculated here for the single NP case. 

\subsection*{Beyond the Mie theory approach}
A more rigorous modal treatment, based on the quasi-normal mode formalism, is possible 
for many of the parameters introduced above \cite{wu_nanoscale_2021}. 
The direct evaluation of $\eta_{\mathrm{rad}}$, named modal brightness $\mathcal{B}$ in the QNM framework, 
remains however a tedious task. 
So, for the contribution of the different loss channels (radiative, absorption, surface plasmon polaritons (SPP)) we followed the approach introduced in Ref.~\citenum{yang_near--far_2016} and evaluated the quenching contribution as described in the section above. The energy dissipated through radiative and guided-mode (SPP) channels could be extinguished by a near-to-far field transformation proposed by Yang et.al \cite{yang_near--far_2016}. The localised absorption could be extracted in a similar way of calculating mode volume in an open cavity proposed by Koenderink \cite{koenderink_use_2010}. By integrating the Ohmic loss density over an increasing domain such that the increasing part (due to the propagating SPPs) could be fitted and removed to find out the intercept that reflects the localized absorption \cite{jiang2020temperature}.
Alternatively, the higher-order modes / quenching contribution could be evaluated by introducing a metal-insulator-metal (MIM) structure equivalent to the structure of interest  \cite{faggiani_quenching_2015,zhang_addressing_2021}. 


\bibliography{bib_appendix}